\documentclass[twocolumn]{aastex631}

\newcommand{\kms}{{\rm km\ s^{-1}}}

\usepackage {threeparttable}
\usepackage{comment}

\begin{document}

\title{Dust Scattering Albedo at Millimeter-Wavelengths in the TW Hya Disk}

\author[0000-0001-8002-8473]{Tomohiro C. Yoshida}
\affiliation{National Astronomical Observatory of Japan, 2-21-1 Osawa, Mitaka, Tokyo 181-8588, Japan}
\affiliation{Department of Astronomical Science, The Graduate University for Advanced Studies, SOKENDAI, 2-21-1 Osawa, Mitaka, Tokyo 181-8588, Japan}

\author[0000-0002-7058-7682]{Hideko Nomura}
\affiliation{National Astronomical Observatory of Japan, 2-21-1 Osawa, Mitaka, Tokyo 181-8588, Japan}
\affiliation{Department of Astronomical Science, The Graduate University for Advanced Studies, SOKENDAI, 2-21-1 Osawa, Mitaka, Tokyo 181-8588, Japan}

\author[0000-0002-6034-2892]{Takashi Tsukagoshi}
\affiliation{Faculty of Engineering, Ashikaga University, Ohmae 268-1, Ashikaga, Tochigi, 326-8558, Japan}

\author[0000-0003-1958-6673]{Kiyoaki Doi}
\affiliation{National Astronomical Observatory of Japan, 2-21-1 Osawa, Mitaka, Tokyo 181-8588, Japan}
\affiliation{Department of Astronomical Science, The Graduate University for Advanced Studies, SOKENDAI, 2-21-1 Osawa, Mitaka, Tokyo 181-8588, Japan}
\affiliation{Max-Planck Institute for Astronomy, Königstuhl 17, D-69117 Heidelberg, Germany}

\author[0000-0002-2026-8157]{Kenji Furuya}
\affiliation{Department of Astronomy, The University of Tokyo, Bunkyo-ku, Tokyo 113-0033, Japan}
\affiliation{RIKEN Cluster for Pioneering Research, 2-1 Hirosawa, Wako-shi, Saitama 351-0198, Japan}

\author[0000-0003-4562-4119]{Akimasa Kataoka}
\affiliation{National Astronomical Observatory of Japan, 2-21-1 Osawa, Mitaka, Tokyo 181-8588, Japan}
\affiliation{Department of Astronomical Science, The Graduate University for Advanced Studies, SOKENDAI, 2-21-1 Osawa, Mitaka, Tokyo 181-8588, Japan}

%% Note that the \and command from previous versions of AASTeX is now
%% depreciated in this version as it is no longer necessary. AASTeX 
%% automatically takes care of all commas and "and"s between authors names.

%% AASTeX 6.31 has the new \collaboration and \nocollaboration commands to
%% provide the collaboration status of a group of authors. These commands 
%% can be used either before or after the list of corresponding authors. The
%% argument for \collaboration is the collaboration identifier. Authors are
%% encouraged to surround collaboration identifiers with ()s. The 
%% \nocollaboration command takes no argument and exists to indicate that
%% the nearby authors are not part of surrounding collaborations.

%% Mark off the abstract in the ``abstract'' environment. 
\begin{abstract}
Planetary bodies are formed by coagulation of solid dust grains in protoplanetary disks.
Therefore, it is crucial to constrain the physical and chemical properties of the dust grains.
In this study, we measure the dust albedo at mm-wavelength, which depends on dust properties at the disk midplane.
Since the albedo and dust temperature are generally degenerate in observed thermal dust emission, it is challenging to determine them simultaneously.
We propose to break this degeneracy by using multiple optically-thin molecular lines as a dust-albedo independent thermometer.
In practice, we employ pressure-broadened CO line wings that provide an exceptionally high signal-to-noise ratio as an optically thin line.
We model the CO $J=2-1$ and $3-2$ spectra observed by the Atacama Large Millimeter/sub-millimeter Array (ALMA) at the inner region ($r<6\ {\rm au}$) of the TW Hya disk and successfully derived the midplane temperature.
Combining multi-band continuum observations, we constrain the albedo spectrum at $0.9-3$ mm for the first time without assuming a dust opacity model.
The albedo at these wavelengths is high, $\sim0.5-0.8$, and broadly consistent with the \citet{ricc10}, DIANA, and DSHARP dust models.
Even without assuming dust composition, we estimate the maximum grain size to be $\sim 340\ \mu m$, power law index of the grain size distribution to be $>-4.1$, and porosity to be $<0.96$.
The derived dust size may suggest efficient fragmentation with the threshold velocity of $\sim 0.08\ {\rm m\ s^{-1}}$.
We also note that the absolute flux uncertainty of $\sim10\%$ ($1\sigma$) is measured and used in the analysis, which is approximately twice the usually assumed value.
\end{abstract}

%% Keywords should appear after the \end{abstract} command. 
%% The AAS Journals now uses Unified Astronomy Thesaurus concepts:
%% https://astrothesaurus.org
%% You will be asked to selected these concepts during the submission process
%% but this old "keyword" functionality is maintained in case authors want
%% to include these concepts in their preprints.

%% From the front matter, we move on to the body of the paper.
%% Sections are demarcated by \section and \subsection, respectively.
%% Observe the use of the LaTeX \label
%% command after the \subsection to give a symbolic KEY to the
%% subsection for cross-referencing in a \ref command.
%% You can use LaTeX's \ref and \label commands to keep track of
%% cross-references to sections, equations, tables, and figures.
%% That way, if you change the order of any elements, LaTeX will
%% automatically renumber them.
%%
%% We recommend that authors also use the natbib \citep
%% and \citet commands to identify citations.  The citations are
%% tied to the reference list via symbolic KEYs. The KEY corresponds
%% to the KEY in the \bibitem in the reference list below. 

\section{Introduction} \label{sec:intro}

The growth of dust grains plays a fundamental role in planet formation.
In the interstellar medium, the maximum dust grain size is thought to be micron meters.
After planet formation was finished, the maximum ``grain'' size reaches $\sim 10^{9}$ cm -- the size of terrestrial planets or planetary cores \citep[e.g.,][]{armi10}.
During the planet formation processes, it is believed that the dust grains grow by coagulation.
However, detailed mechanisms of grain evolution are still unclear.
To reveal these processes, it is necessary to characterize the dust grains in protoplanetary disks, where the dust grains experience the most dramatic evolution.

The physical and chemical characteristics of the dust grains, such as the composition, size distribution, and structure, affect the optical properties.
Therefore, observational constraints on such properties are significantly worthwhile.
Indeed, infrared observations can give some constraints since some minerals and ice have broad features in their spectra \citep[e.g.,][]{ober21}.
However, infrared wavelengths usually trace the elevated layer of protoplanetary disks \citep[e.g.,][]{aven18}, where small dust particles dominate.
Since planet formation { is expected to occur} at the midplane, observations with longer wavelengths would be helpful although it is known that the spectra in the longer wavelengths are feature-less.

In this paper, we particularly focus on the scattering albedo at the mm-wavelengths.
%Recent observational studies suggest that dust scattering affects its emission significantly even in the mm-wavelengths.
The albedo as a function of wavelength (hereafter {\it albedo spectrum}) depends on the dust property, and therefore, would give us fruitful information without a feature as seen in infrared wavelengths.
However, it is challenging to directly constrain the albedo spectrum by observations.
Although some previous studies of the dust continuum emission suggest that the albedo is high \citep[$\sim0.8-0.9$; e.g., ][]{carr19, ueda20}, they assume some kind of dust models.
For example, \citet{carr19} assumed that the albedo follows a power law.
\citet{ueda20} adopted the DSHARP dust model \citep{birn18} which has intrinsically high albedo.

Indeed, it is impossible to model-independently derive the albedo by using only continuum observations.
It is known that scattering induces the intensity reduction in optically thick regime  \citep{miya93, zhu19, ueda20}.
We see the effect of the scattering-induced intensity reduction in the following.
The emission from a homogeneous optically thick dust slab is generally expressed as
\begin{equation}
\label{eq:eq1}
    I = \chi(\omega) B (T),
\end{equation}
where $\chi(\omega)$ is the intensity reduction factor at the albedo $\omega$, and $B(T)$ is the Planck function and the temperature $T$ \citep{zhu19}.
When $\omega \sim 1$, $\chi(\omega)$ is approximated to be $\sqrt{ 1-\omega}$.
Therefore, the albedo generally degenerates with temperature; we cannot distinguish between an increase in temperature and a decrease in the albedo.

Meanwhile, it would be possible to independently constrain the temperature using molecular emission lines as the dust and gas have the same temperature in relatively high-density regions.
Indeed, so-called rotation diagrams of optically thin molecular emission lines are routinely used to measure the temperature (and the column density) in molecular clouds as well as disks \citep[e.g., ][]{yama17}.
However, the situation is more challenging near the midplane of protoplanetary disks.
This is essentially due to the lack of molecular species that trace near the midplane with a sufficiently high signal-to-noise ratio.

Recently, \citet{yosh22} reported a pressure-broadened line wing of the CO emission spectrum in the inner region of the closest protoplanetary disk around TW Hya \citep[$D\sim 60.1$ pc;][]{gaia16, gaia21}.
Since the pressure-broadened line wings are optically thin and essentially trace near the disk midplane inside the CO snowline, we can use it as an independent thermometer with a relatively high signal-to-noise ratio if two or more CO transitions are available.
The constraints on the temperature and the intensity of optically thick emission give the intensity reduction factor, using Equation (\ref{eq:eq1}), which can be then converted to the albedo.

In this paper, we first describe a conceptual formulation of our strategy in Section \ref{sec:method}.
In the subsequent sections, we focus on the inner region ($R \lesssim 10$ au) of the protoplanetary disk around TW Hya and constrain the dust albedo spectrum.
\begin{figure}[hbtp]
    \epsscale{1.1}
    \plotone{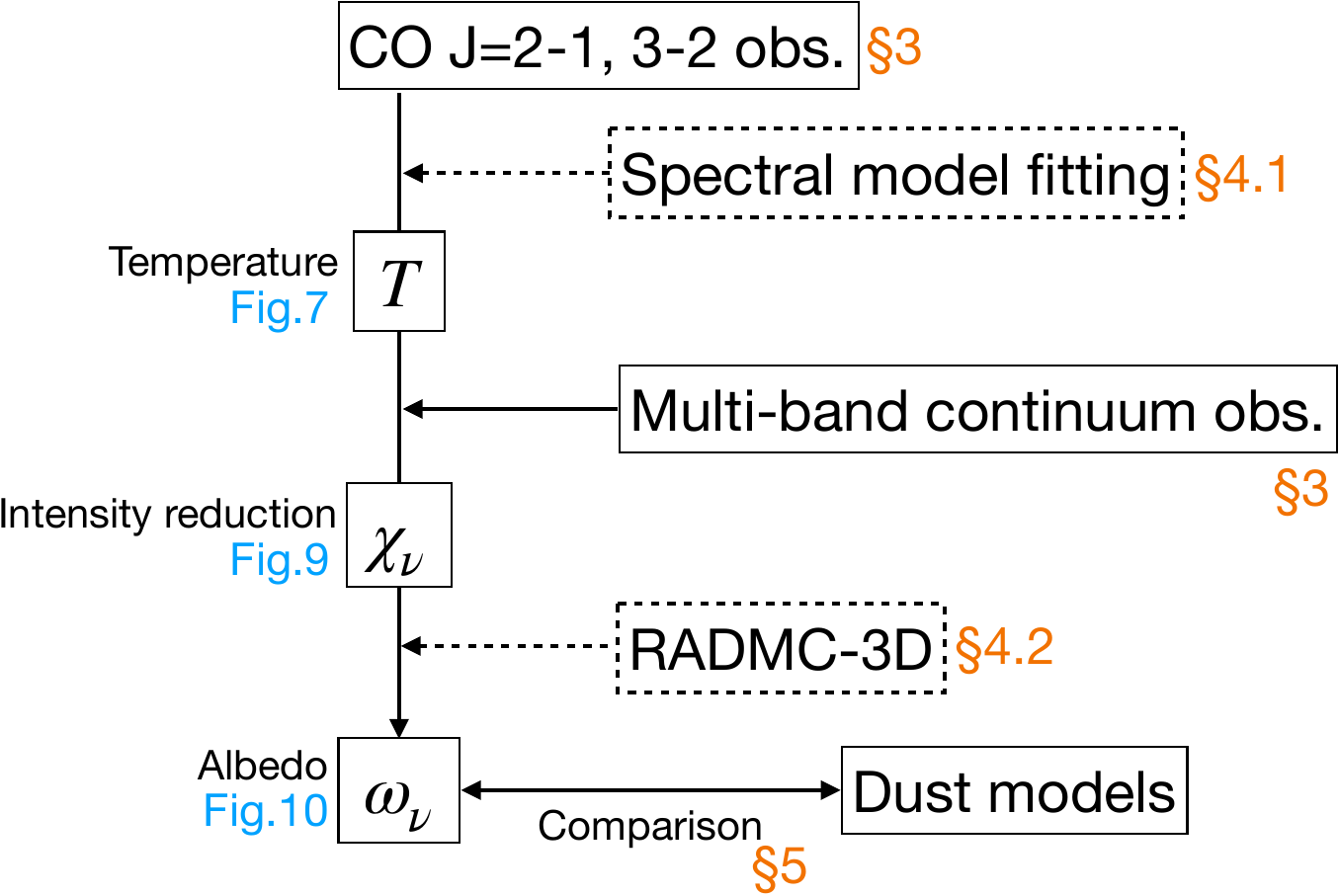}
    \caption{Flowchart of the procedure of this paper. }
    \label{fig:ponchi}
\end{figure}
The procedure after Section \ref{sec:obs} is summarized in Figure \ref{fig:ponchi}.
In Section \ref{sec:obs}, archival observations that we used in this paper are briefly introduced.
In Section \ref{sec:midtemp}, we construct a simple model, fit it to the observed spectra of CO $J=2-1$ and CO $J=3-2$ lines, and derive the temperature.
Then, by comparing the best-fit temperature model and the continuum spectral energy distribution (SED), we constrain the albedo spectrum at 0.9 - 3 mm wavelengths.
The obtained albedo spectrum is further compared with { the Ricci \citep{ricc10}}, DIANA \citep{woit16} and DSHARP \citep{birn18} dust models in Section \ref{sec:dustmodels}.
Furthermore, we free the parameters on composition in the DSHARP model and fit the albedo spectrum to constrain the size distribution and porosity.
We discuss the results in Section \ref{sec:disc} and conclude this paper in Section \ref{sec:conc}.

\section{Concept of the Method} \label{sec:method}
In this section, we describe the concept of our method to observationally determine the albedo spectrum.
%We note that the results of this section are not directly applied to observations in the later section.
{ Here, we aim to show that the albedo spectrum can be obtained independently of dust models by a combination of dust continuum and molecular line emission.
We introduce a more practical methodology for application to observations in the later sections.
}

\subsection{Line Emission with Continuum Background}

\begin{figure}[hbtp]
    \epsscale{0.7}
    \plotone{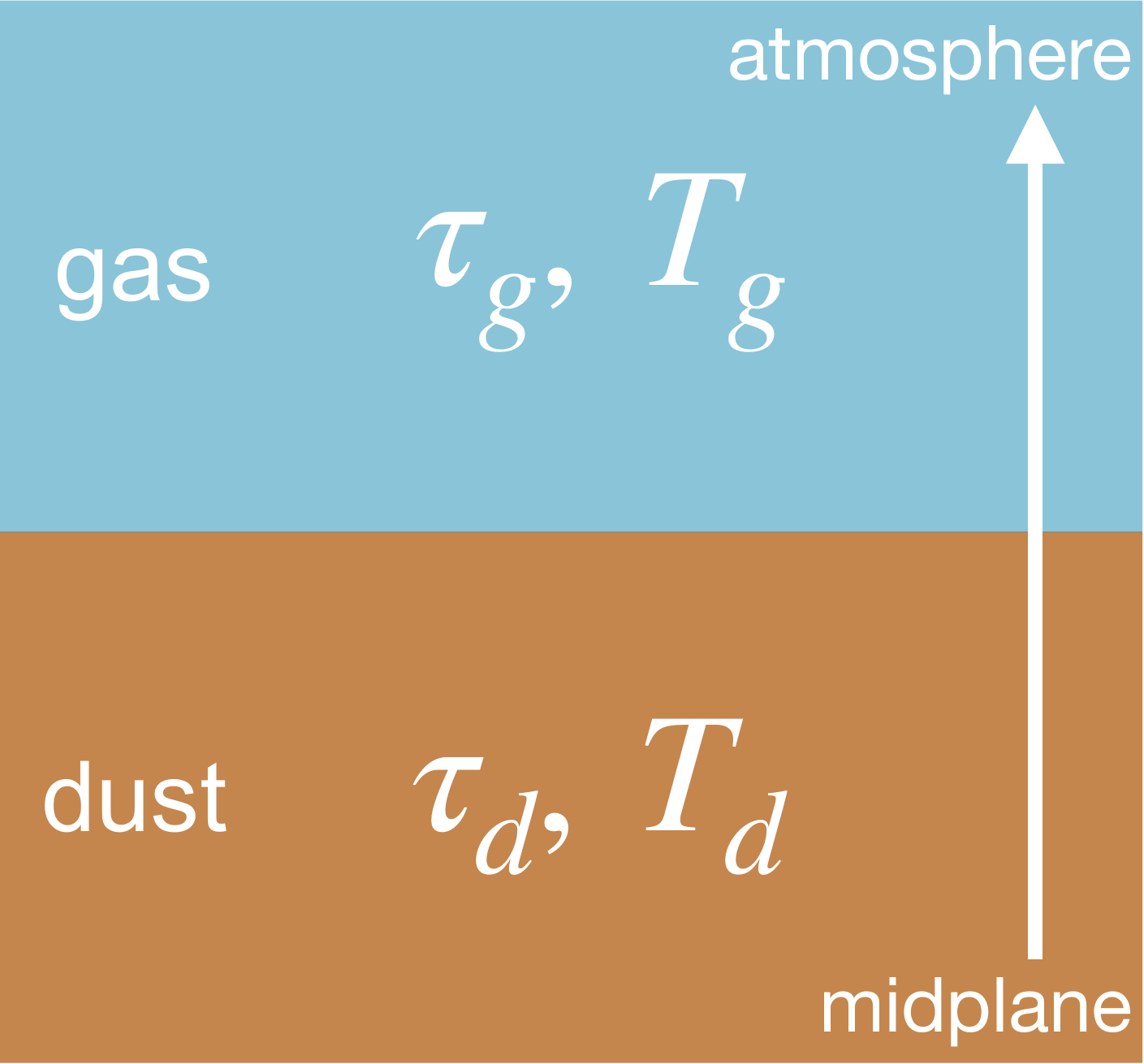}
    \caption{ Layered structure discussed as a toy model. The white arrow indicates the propagation direction of a ray toward the observer.}
    \label{fig:toymodel}
\end{figure}

First, we employ a toy model to illustrate the method.
We assume a layered one-dimensional homogeneous dust and molecular gas structure as shown in Figure \ref{fig:toymodel}, which is similar to a toy model considered by \citet{bosm21}.
Here, the optical depths of the gas and dust layers are $\tau_g$ and $\tau_d$, respectively. Similarly, the gas and dust temperatures are $T_g$ and $T_d$, respectively.
Considering radiative transfer along the direction from dust to gas layers, the emerging intensity $I_t$ can be expressed as
\begin{equation}
    I_t = \chi_d(\tau_d, \omega) B(T_d) ( 1 - e^{-\tau_d} ) e^{-\tau_g} + B(T_g) ( 1 - e^{-\tau_g} ),
\end{equation}
where $\chi_d(\tau_d, \omega)$ is the intensity reduction factor due to scattering.
When the dust emission is optically thick and the albedo is high, $\chi_d(\tau_d, \omega) = \chi(\omega)$.
If we assume the face-on geometry, $\chi(\omega)$ can be approximated as
\begin{equation}
    \chi(\omega) = \frac{2 + \sqrt{3}}{ 2 + \frac{\sqrt{3}}{ \sqrt{1-\omega} } } \label{eq:chi},
\end{equation}
as shown in \citet{zhu19}.
Then, the intensity becomes
\begin{equation}
    I_t = \chi(\omega) B(T_d) e^{-\tau_g} + B(T_g) ( 1 - e^{-\tau_g} )
\end{equation}
at the limit of optically thick dust emission.
Here, we consider the case that the molecular gas emission is { very} optically thin, $\tau_g \ll 1$:
\begin{equation}
    I_t = \chi(\omega) B(T_d) (1-\tau_g) + B(T_g) \tau_g. \label{eq:It0}
\end{equation}
In spectroscopic observations, we can observe the continuum emission without a molecular line, that is $\chi(\omega) B(T_d)$.
As a result, the continuum-normalized intensity $i_t$ can be written as
\begin{eqnarray}
    i_t &=& \frac{I_t}{\chi(\omega) B(T_d)} \\
    &=&  (1-\tau_g) +  \frac{1}{\chi(\omega)} \frac{B(T_g)}{B(T_d)} \tau_g, \\
    &=&  1 + \left( \frac{1}{\chi(\omega)} \frac{B(T_g)}{ B(T_d)} - 1 \right) \tau_g \\
    &\equiv& 1 + f_g \tau_g. \label{eq:it}
\end{eqnarray}
Here, we define $f_g$ as a line emerging factor per a unit line optical depth.
$f_g$ depends on both $B(T_g)/B(T_d)$ and $\omega$ in general.
In the case we treat in this paper, however, we can reasonably assume $B(T_g)/B(T_d)=1$, and therefore, $f_g$ only depends on $\omega$.
We discuss this assumption in Section \ref{sec:self_consistency}.

\begin{comment}

We plot $f_g$ as a function of $\omega$ and $B(T_g)/B(T_d)$ in Figure \ref{fig:fg}.
\begin{figure}[hbtp]
    \epsscale{1}
    \plotone{./figure/fg.pdf}
    \caption{ Line emerging factor $f_g$ as a function of $\omega$ and $B(T_g)/B(T_d)$.}
    \label{fig:fg} 
\end{figure}
This figure shows that we can observe optically thin line emission above optically thick continuum emission only when the albedo is high and/or the vertical temperature gradient is large.

For example, if $\omega=0.8$ and $B(T_g)/B(T_d) = 1$, the line emerging factor $f_g$ is $\sim 0.6$.
To maintain the same intensity for fixed line optical depth without scattering,  $B(T_g)/B(T_d) \sim 1.6$ is needed.

In protoplanetary disks, the temperature should be almost constant along the vertical axis within one or two gas scale heights from the midplane \citep{dull02, pint09, inou09}.
Additionally, the dust and gas are thermally coupled when the density is sufficiently high.
Therefore, under the assumption that a molecular line originates near the midplane, $f_g$ only depends on the albedo $\omega$. 
%Figure \ref{fig:fg2} illustrates $f_g$ as a function of $\omega$ when $T_g = T_d$.
In other words, if we can constrain $\tau_g$ independently of the continuum emission, the albedo can be determined from observations.
\end{comment}

\subsection{Deriving Temperature and Albedo} \label{sec:2.2}
In this subsection, we describe a formulation to derive the temperature and albedo simultaneously.
Hereafter, we assume (1) the dust and line emitting regions are layered (Figure. \ref{fig:toymodel}; see also Section.\ref{sec:self_consistency}), and (2) temperature is vertically constant within the emitting layer of gas and dust ($T_g = T_d$).
{ In practice, we can choose emission lines that satisfy these assumptions as we explain later.}
%As shown in the following, the gas and dust temperature can be determined independently of the albedo by using two transitions of the same molecule under some assumptions.

The line optical depth $\tau_g$ only depends on the transition, temperature, and column density assuming the local thermal equilibrium (LTE) \citep{rybi79}.
If we observe two transition lines of the same molecule and denote them by subscripts 1 and 2, the observed quantities before continuum normalization are
\begin{eqnarray}
    I_{t, 1} &=& \chi(\omega_1) B_1(T_d) + ( 1 -  \chi(\omega_1)) B_1(T_d) \tau_{g, 1},  \label{eq:It1} \\
    I_{t, 2} &=& \chi(\omega_2) B_2(T_d) + ( 1 -  \chi(\omega_2)) B_2(T_d) \tau_{g, 2},\label{eq:It2}
\end{eqnarray}
assuming $T_g = T_d$.
Simultaneously, we can obtain the continuum emissions as
\begin{eqnarray}
\label{eq:cont_toymodel}
    I_{c, 1} &=& \chi(\omega_1) B_1(T_d), \label{eq:c1} \\
    I_{c, 2} &=& \chi(\omega_2) B_2(T_d). \label{eq:c2}
\end{eqnarray}
Under the LTE condition, the ratio of the line optical depths only depends on the temperature.
Therefore, we can express
\begin{equation}
    \frac{\tau_{g, 1}}{\tau_{g, 2}} = C(T_d),  \label{eq:tau}
\end{equation}
by defining a function $C(T_d)$ that only depends on $T_d$ and known spectroscopic, physical constants.

We can solve the Equations (\ref{eq:It1})-(\ref{eq:tau}) for $\chi(\omega_1)$, $\chi(\omega_2)$, and $T_d$ simultaneously.
First, we substitute $\chi(\omega_1) = {I_{c, 1}}/{B_1(T_d)}$ and  $\chi(\omega_2) = {I_{c, 2}}/{B_2(T_d)}$ to Equations (\ref{eq:It1}) and (\ref{eq:It2}).
Then, we solve Equations (\ref{eq:It1}) and (\ref{eq:It2}) for $\tau_{g, 1}$ and $\tau_{g, 2}$, take their ratio, equate it by Equation (\ref{eq:tau}), and obtain
\begin{equation}
    \frac{B_2(T_d) - I_{c,2}}{B_1(T_d) - I_{c,1}} \cdot \frac{I_{t, 1} - I_{c,1}}{I_{t, 2} - I_{c, 2}} = C(T_d). \label{eq:T}
\end{equation}
It is possible to implicitly solve Equation (\ref{eq:T}) for $T_d$.
Once $T_d$ is determined, we can substitute it to Equations (\ref{eq:c1}) and (\ref{eq:c2}) and get $\chi(\omega_1)$ and $\chi(\omega_2)$.
Finally, we can constrain the albedo $\omega_1$ and $\omega_2$ via Equation (\ref{eq:chi}).
In addition, if more continuum observations at different wavelengths are available, the albedo $\omega$ at each wavelength can be determined using the temperature $T_d$ obtained from the line analysis via equations equivalent to Equation (\ref{eq:cont_toymodel}).

To apply this method in practice, we have to obtain a sufficient signal-to-noise ratio for the optically thin emission. 
%For instance, if $\tau_g = 0.1$, the continuum-normalized intensity is only $\sim 1.057$ and $\sim 1.019$ for $\omega = 0.8$ and 0.5, respectively, which means that we need to measure the continuum-normalized line intensity with $< 1 \%$ accuracy.
%If the continuum emission has a brightness temperature of 30 K, we have to measure the line flux with 0.3 K accuracy.
%This is not impossible with the Atacama Large Millimeter/sub-millimeter Array, but it takes $\sim 10$ hours integration in Band 7 according to the Sensitivity Calculator.
Furthermore, it is required to guarantee that the emission arises near the midplane.
One of the best candidates that satisfy the above conditions is the pressure-broadened CO line wings \citep{yosh22}.
\citet{yosh22} found that the CO $J=3-2$ line in the TW Hya disk has very broad line wings spanning from $-10\ \kms$ to $+10\ \kms$ with respect to the line center and attributed it to the pressure broadening \citep[][Sec 10.6]{rybi79}.
The broad line wings are optically thin but have a width of $\sim 10\ \kms$, which is exceptional as optically thin line emission.

%When the albedo $\omega$ is $0.8$ and $T_g = T_d$, the continuum-normalized intensity is $1 + 0.57 \tau_g$ for instance. Assuming the line optical depth of $\tau_g \sim 0.1$, the line emission can be observed as an increase by $\sim 5 \%$ from the continuum level.
%On the other hand, if $\omega = 0$, $B(T_g)$ has to be $1.57 \times B(T_d)$ to observe the same level of line emission.

\section{Archival Observations} \label{sec:obs}
In this section, we introduce observational data that is used in subsequent analysis. 
{ We analyze the CO $J=2-1$ and $3-2$ lines as well as the continuum emission at ALMA Band 3 ($\lambda = 3.2\ {\rm mm}$), 4 ($2.1\ {\rm mm}$), 6 ($1.2\ {\rm mm}$), 7 ($0.87\ {\rm mm}$), and 8 ($0.63\ {\rm mm}$) in the TW Hya disk.}
All the images are re-convolved with a 2D-gaussian to have an elliptical beam with an FWHM of 12 au (0\farcs2), corresponding to a disk radius of 6 au, when the images are deprojected so that the disk becomes in a face-on view.
Throughout the following analysis, we adopt the disk inclination angle of $5\fdg8$ and position angle of $242^\circ$ \citep{teag19}.

\subsection{CO Lines}

For the CO $J=3-2$ line, we used the same data presented in \citet{yosh22} although we did not subtract the continuum emission.
The image cube was created from ALMA archival observations. The project IDs included in this data are 2015.1.00686.S (PI: S.Andrews), 2016.1.00629.S (PI: I. Cleeves), and 2018.1.00980.S (PI: R. Teague).

We also used the image cube of the CO $J=2-1$ line, which is generated from ALMA archival observations, 2018.A.00021.S (PI: R. Teague).
The observations were described in \citet{cant21} and \citet{teag22}.
The data we used was based on the self-calibrated measurement sets presented in \citet{yosh24} who focuses on the CN lines observed simultaneously.
We CLEANed the CO $J=2-1$ line with a robust parameter of 0.5 with a channel width of $0.25\ \kms$.

\begin{figure*}[hbtp]
    \epsscale{1}
    \plotone{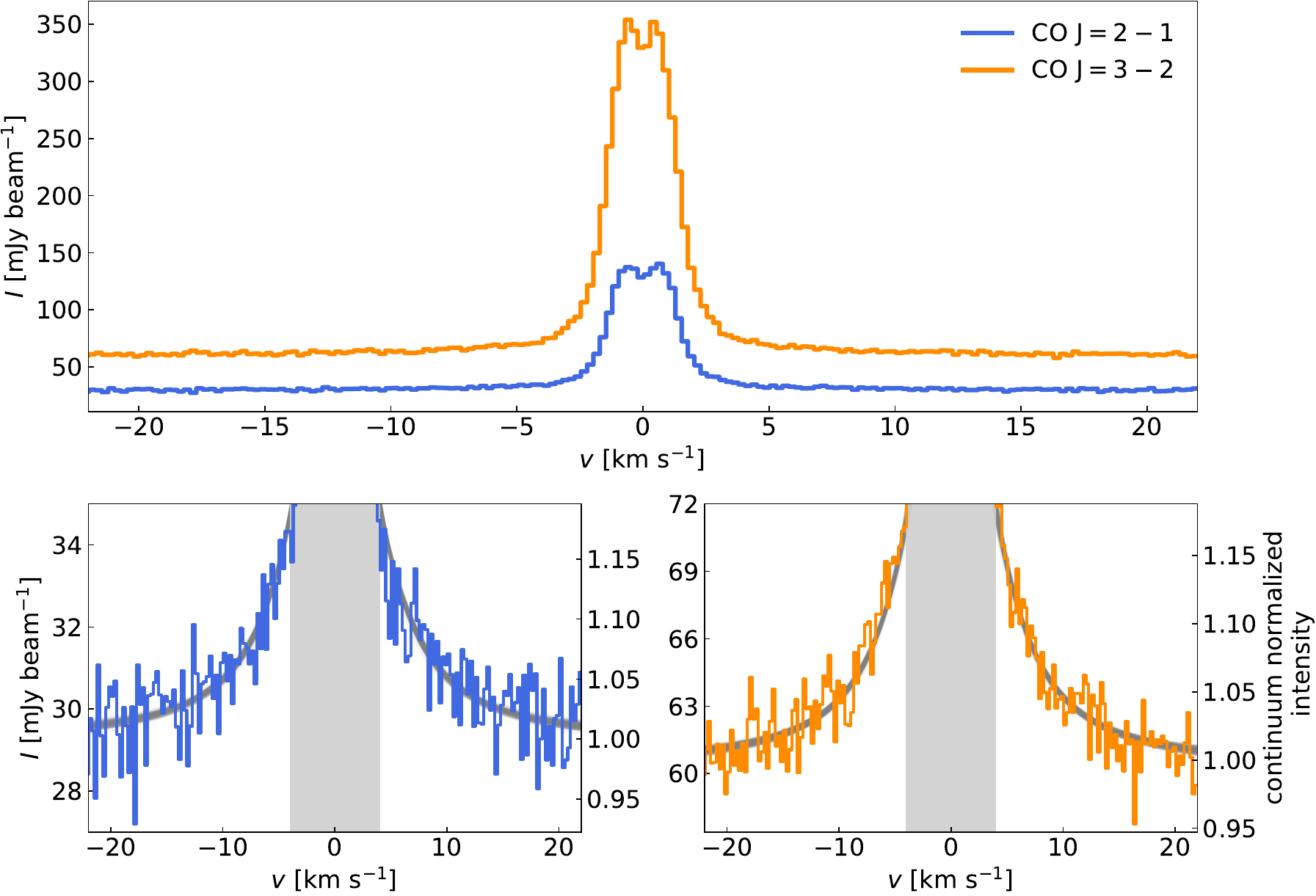}
    \caption{ CO $J=2-1$ (blue) and $J=3-2$ (orange) spectra at the center ($r<6\ {\rm au}$) of the TW Hya disk. The bottom panels are the zoom-in version of the top panel. The grey lines indicate the models created from parameters randomly sampled from the MCMC chain (Section \ref{sec:res1}). The gray-shaded regions ($|v| < 4\ {\rm km\ s^{-1}}$) are not used for fitting.}
    \label{fig:spec} 
\end{figure*}

The CO spectra at the center of the disk after re-convolution are plotted in Figure \ref{fig:spec}.
Note that the centers of the disk were specified by measuring the emission barycenter at the far-wing velocities.
The rms noise levels of the two extracted spectra are 0.55 ${\rm mJy\ beam^{-1}}$ and 0.94 ${\rm mJy\ beam^{-1}}$ for the CO $J=2-1$ and $J=3-2$ lines, respectively.

\subsection{Continuum}

For the Band 6 and 7 continuum images, we used the corresponding continuum images for the CO line data.
The Band 3 and 4 images presented in \citet{tsuk22} are also used.

The Band 8 image are newly analyzed.
We obtained the visibility data sets from the ALMA science archive (2015.1.01137.S, PI: T. Tsukagoshi; 2016.1.01399.S, PI: V. Salinas).
First, the visibilities are calibrated by the pipeline.
Then, we executed four rounds of phase-only self-calibration for the shorter baseline continuum data (2016.1.01399.S) with the shortest solution interval of 60 seconds.
The self-calibrated data was concatenated with the longer baseline data, and we ran one round of phase and amplitude self-calibration with an interval of the execution block duration.
The continuum image is generated by the CLEAN algorithm with a robust parameter of $-1.5$.
The final noise level is 0.65 $\rm mJy\ beam^{-1}$.

%The Band 9 images were created from archival observations of 2012.1.00422.S (PI: E. Bergin). (details?)
%However, it was not possible to obtain the 0\farcs2 beam size by CLEAN.
%Therefore, we tried to make an image with a imaging method based on sparse modeling \citep{yama20}.
%Details of the imaging process are described in Appendix A.

In addition, to assess the absolute flux uncertainty and re-scale absolute fluxes, we collected almost all product images in the science archive.
We explain the detailed procedure in Appendix \ref{sec:fluxscaling}.
Overall, we found that the uncertainty of product images is $\sim2-4$ times larger than the nominal flux uncertainty, which is consistent with \citet{fran20}.
However, thanks to a large number of independent archival observations, we can reduce the uncertainty by a factor of $\sim 2-5$ (see Appendix \ref{sec:fluxscaling} for details).

\begin{figure}[hbtp]
    \epsscale{1.0}
    \plotone{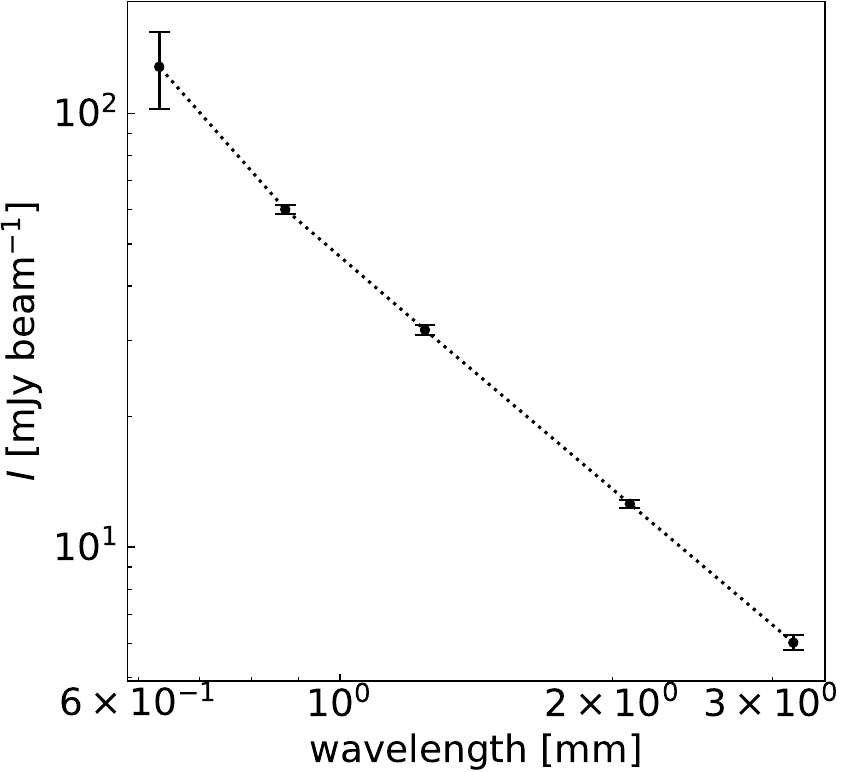}
    \caption{ Continuum SED at the center ($r<6\ {\rm au}$) of the disk. { The uncertainty of Band 3 ($\lambda = 3.2\ {\rm mm}$), 4 ($2.1\ {\rm mm}$), 6 ($1.2\ {\rm mm}$), and 7 ($0.87\ {\rm mm}$) fluxes are estimated from archival images (Appendix \ref{sec:fluxscaling}) while that of Band 8 ($0.63\ {\rm mm}$) is a nominal but conservative value, $\sim20 \%$.}}
    \label{fig:sed} 
\end{figure}

The re-scaled SED extracted at the center pixel of the images is shown in Figure \ref{fig:sed}.
The error bars of Band 3-7 are based on the analysis of product images, that is,  3.9, 2.2, 2.7, and 2.5\% for Bands 3, 4, 6, and 7, respectively.
Note that these uncertainties are much larger than the thermal noise levels.
On the other hand, the uncertainty of the Band 8 flux is conservatively assumed to be $\sigma \sim 20\ \%$, which is stated as a worse case in the ALMA technical handbook.

\section{Constraining the Midplane Temperature and Albedo Spectrum} \label{sec:midtemp}

\subsection{Emission Model Fitting} \label{sec:emimod}
Although we already showed that the albedo spectrum can be observationally determined using a slab model, it is needed to properly treat the disk structure as well as the gas kinematics for accurate measurements.
Therefore, we take a forward modeling approach and fit an emission model to the spectra extracted at the TW Hya disk center (Figure \ref{fig:spec}).
We describe our assumptions and details of the model in the following.

The midplane temperature structure $T(r)$ is assumed to follow a power law of the radius from the central star $r$,
\begin{equation}
    T = T_0  \left ( \frac{r}{r_0} \right )^{-0.5}.
\end{equation}
$T_0$ is the temperature at $r=r_0$, where $r_0$ is the radius of the inner dust cavity seen in a high spatial resolution image \citep{andr16}.
While we fixed $r_0 = 3.2$ au according to \citet{yosh22}, we treat $T_0$ as a free parameter.
The power law index of $-0.5$ is motivated by the multi-band continuum modeling results of \citet{maci21}.

\citet{maci21} also suggested that the dust continuum emission is highly optically thick even in Band 3 at $r \lesssim 20$ au, where the pressure-broadened CO line wings arise from \citep{yosh22}.
Therefore, we assume that the dust continuum emission outside the dust cavity at frequency $\nu$, $I_{d, \nu}$, can be expressed as 
\begin{equation}
    I_{d, \nu} = \chi_\nu B_\nu (T),
\end{equation}
where $\chi_\nu$ is the intensity reduction factor due to scattering and $B_\nu (T)$ is the Planck function at the frequency $\nu$ and temperature $T$.
$I_{d, \nu}$ inside the cavity is the same as the cosmic background radiation with a black body at 2.73 K.
Note that this corresponds to Equation (\ref{eq:cont_toymodel}) in the toymodel.

We also assume that the gas surface density structure follows a power law.
The gas surface density profile $\Sigma_g$ is assumed to be
\begin{equation}
    \Sigma_g = \Sigma_{g,0}  \left ( \frac{r}{r_0} \right )^{-0.5},
\end{equation}
where $\Sigma_{g,0}$ is the gas surface density at the cavity radius.
This is also treated as a free parameter.
Meanwhile, we fixed the radial structure of the gas surface density as our analysis in this paper does not have spatial information.
We adopted the observed power law index of 0.5 by following \citet{yosh22} which uses the visibility data itself.
Inside the cavity, the gas surface density is depleted by a factor of $\simeq42$ \citep{yosh22}.
The CO column density $N_{\rm CO}$ is then expressed as 
\begin{equation}
    N_{\rm CO} = \frac{ \Sigma_g}{ 2 \mu_g m_p } X_{\rm CO}.
\end{equation}
Here, $\mu_g$, $m_p$, and $X_{\rm CO}$ are the mean molecular weight of 2.37, proton mass, and CO/H$_2$ density ratio, respectively.
$X_{\rm CO}$ is also kept as a free parameter.
The $\rm H_2$ gas volume density $n_{\rm H_2}$ that is needed to reproduce the pressure-broadened line wings is calculated by
\begin{equation}
    n_{\rm H_2} = \frac{ \Sigma_g}{  2 \sqrt{\pi} \mu_g m_p H_g }.
\end{equation}
$H_g$ is the gas scale height and expressed as
\begin{equation}
    H_g = \sqrt{ \frac{ k_B T r^3 }{ G M \mu_g m_p } },
\end{equation}
assuming the vertical hydrostatic equilibrium, where $k_B$, $G$, and $M$ are the Boltzmann constant, gravitational constant, and central stellar mass of $0.81\ M_\odot$ \citep{teag19}.

Using the above quantities, the spectra of CO J=2-1 and 3-2 at each radius in the disk can be calculated.
We adopted the Voigt function as a line profile \citep[][Sec. 10.6]{rybi79} and obtained the optical depths of the two CO lines, $\tau_i(v)$, at each radius from 0.01 au to 32 au. Here, the subscript means CO $J=2-1$ or $3-2$, and $v$ is the velocity from the line center.
The detailed formulation followed \citet{yosh22}. We used the spectroscopic parameters in the LAMDA database \citep{lamda} and HITRAN database \citep{hitran}.
As a result, the spectra as a function of radius are
\begin{equation}
    I_i(r) = I_{d, \nu_i} e^{-\tau_i(v)} + B_{\nu_i} (T) ( 1-e^{-\tau_i(v)} ) \label{eq:singlespec},
\end{equation}
where $\nu_i$ indicates the frequency at each transition.
To obtain the spectra at the center of the disk, we followed \citet{bosm21}.
First, we convolved Equation (\ref{eq:singlespec}) along $v$ by
\begin{equation}
    \frac{1}{\sqrt{ v_{\rm k}^2 - v^2  }},
\end{equation}
where $v_{\rm k}$ is the maximum Keplerian velocity at the radius $r$,
\begin{equation}
    v_{\rm k} = \sqrt{\frac{GM}{r}} \sin{ \theta_i }.
\end{equation}
$5\fdg 8$ is adopted for the inclination angle $\theta_i$.
Then, the radially-dependent spectra are integrated over the radius with a mask that expresses the beam sensitivity distribution.

In summary, we have five free parameters in our model; $\Sigma_{g,0}$, $X_{\rm CO}$, $T_0$, and two intensity reduction factors $\chi_\nu = \chi_6, \chi_7$ at frequencies of CO $J=2-1$ or $3-2$.
The generated spectra were compared with the observations in terms of the chi-squared.
Here, to prevent contamination from the optically thick CO emission near the line center, we masked $|v| < 4\ \kms$ and only used the velocity ranges of $|v| > 4\ \kms$ (Figure \ref{fig:spec}).
The posterior distribution of the log-likelihood function was sampled using the Markov-Chain Monte-Carlo (MCMC) method with a publicly available code {\tt emcee} \citep{emcee}.
Our MCMC chain consists of 10,000 steps with 20 walkers.
We discarded the first 5,000 steps as burn-in samples.
Uniform priors are adopted with broad boundaries listed in Table \ref{tab:prior}.
\begin{table}[hbtp]
  \caption{Priors adopted for the MCMC fitting}
  \label{tab:prior}
  \centering
  \begin{tabular}{ccc}
    \hline
    Parameter & Range & Unit \\
    \hline \hline
    $\log_{10} \Sigma_{g,0}$  & $[1, 5]$  & $\rm g\ cm^{-2}$ \\
    $\log_{10} X_{\rm CO}$  & $[-7, -4]$   & \\
    $T_0$  & $[1, 2000]$ & $\rm K$ \\
    $\chi_6, \chi_7$  &  $[0, 1]$  &   \\
    \hline
  \end{tabular}
\end{table}
After determining the temperature $T$, we can derive the intensity reduction factors $\chi_\nu$ at each wavelength available in the continuum observations by fixing $\tau_i(v) = 0$ in the above equations.

\subsection{$\chi-\omega_{\rm eff}$ relation based on RADMC-3D models}
The intensity reduction factor $\chi$ can be then converted to the dust effective scattering albedo as described below.
The effective scattering albedo is defined by
\begin{equation}
\label{eq:sca_opa}
 \omega_{\rm eff} = \frac{\kappa_s^{\rm eff}}{\kappa_a+\kappa_s^{\rm eff}},
\end{equation}
where $\kappa_s^{\rm eff}$ and $\kappa_a$ are the effective scattering opacity and absorption opacity.
The effective scattering opacity includes the anisotropic scattering effect to some extent as discussed in Section \ref{sec:dustmodels}. 
According to \citet{zhu19}, the intensity reduction factor is in a one-to-one correspondence if the continuum emission is optically thick, and the disk geometry is fixed.
This $\chi-\omega_{\rm eff}$ relation could be approximated by analytical formulae such as Equation \ref{eq:chi}.
However, they also pointed out that the analytical formulae are somewhat different from the results of numerical simulations.
Therefore, we choose to calculate this relation using a 3D Monte-Carlo radiative transfer code {\tt RADMC-3D} \citep{dull12}.

Our dust disk model for the simulation is constructed based on the above CO emission model in Section \ref{sec:emimod}.
We assume that the dust surface density profile is $0.01\times \Sigma_g$ and the dust scale height is $0.05\times H_g$.
%$\Sigma_g$ is not well constrained by the above emission model fitting because of absence of optically thin CO isopologue line in this paper as mentioned in the next sub-section.
{ Here, we adopt $\Sigma_{g, 0} = 1500\ {\rm g\ cm^{-2}}$, which is measured through the pressure-broadened CO line wings and optically thin $\rm ^{13}C^{18}O$ line \citep{yosh22}.
We note that an optically thin isotopologue line is needed to determine the gas surface density, but this is irrelevant to this study (Section \ref{sec:res1}).
For $H_g$, we substituted the best-fit temperature value of the emission model fitting.}
The inclination angle of $5.8^\circ$ is also adopted.
The dust absorption opacity is fixed to be $\kappa_a = 2.7\ {\rm cm^2\ g^{-1}}$, which ensures that the dust disk is highly optically thick, $\tau_d \simeq 21$, even at the edge of the focused region ($12$ au).
We note that the resulting $\chi - \omega_{\rm eff}$ relation does not depend on $\kappa_a$ as long as the emission is optically thick.

We calculate the continuum flux density with and without scattering while changing the effective scattering opacity $\kappa_s^{\rm eff}$.
In the RADMC-3D calculation, we use the isotropic scattering mode ({\tt scattering\_mode} $= 1$) for the case with scattering.
Note that we introduce the anisotropic scattering effect with the forward-scattering parameter when we compare the observed albedo with various dust models (see Section \ref{sec:dustmodels}).
Additionally, we set 256 pixels over the calculation domain of 32 au and employ $10^{6}$ photons for Monte Carlo calculations.
We ran the simulations at the wavelength of 1 mm in practice, but the results do not depend on the choice of wavelength.

\begin{figure}[hbtp]
    \epsscale{1.0}
    \plotone{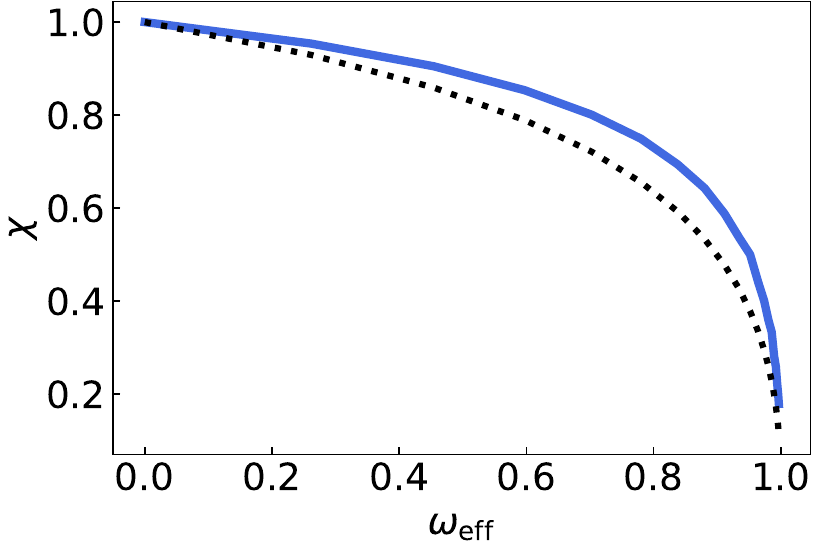}
    \caption{ Intensity reduction factor as a function of the effective albedo. The blue line is the results from the numerical simulation while the black dotted line indicates an approximated formula in Equation (\ref{eq:chi}). }
    \label{fig:chiomega} 
\end{figure}

The ratio of the flux density with scattering to one without scattering is the intensity reduction factor.
Figure \ref{fig:chiomega} shows the calculated intensity reduction factor as a function of the effective albedo.
As a reference, we showed the approximated form in Equation (\ref{eq:chi}), which have generally lower $\chi$ compared to the derived values here.

\subsection{Results} \label{sec:res1}

\begin{figure*}[hbtp]
    \epsscale{1.0}
    \plotone{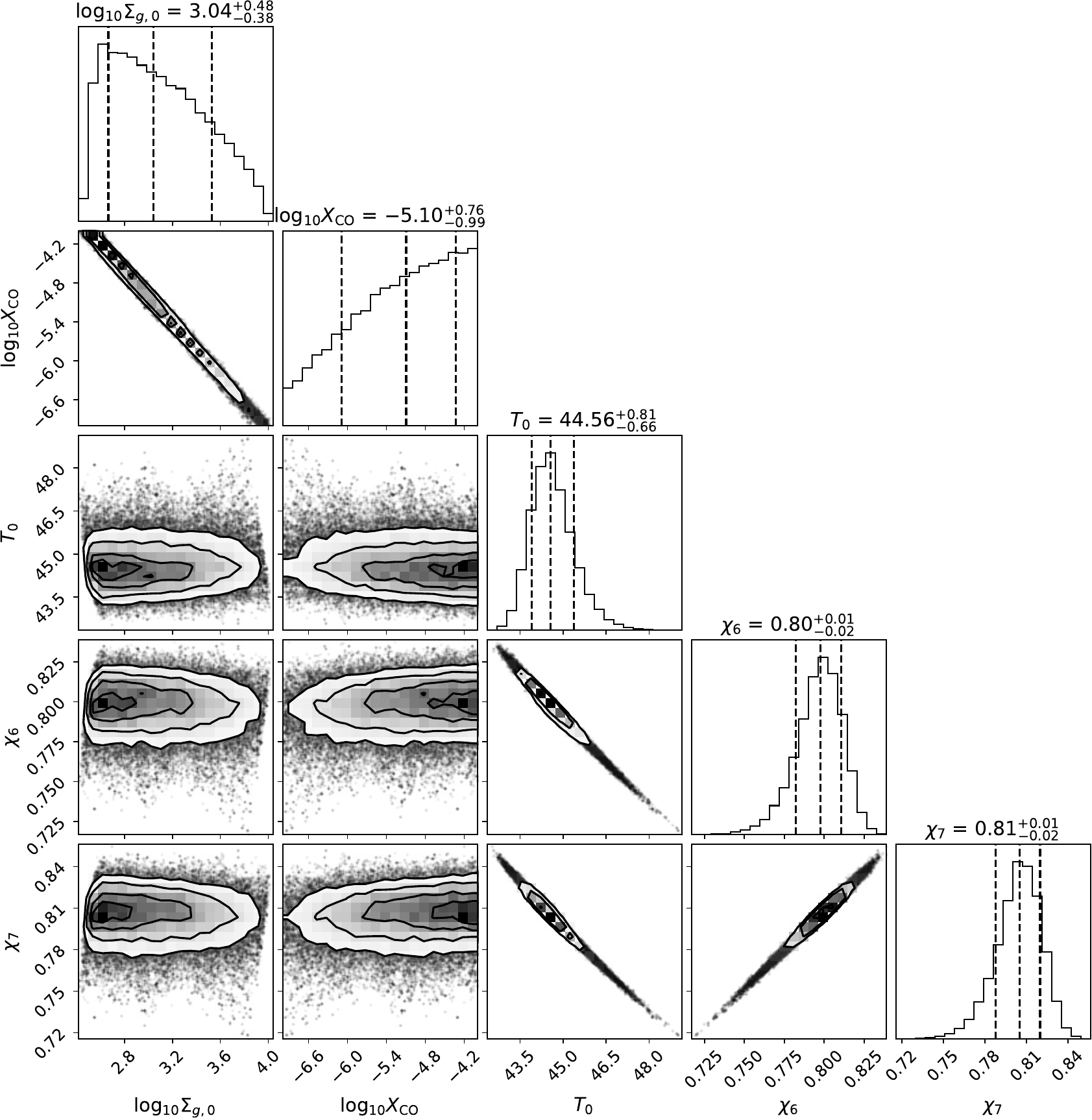}
    \caption{ Corner plot of the MCMC fitting of the emission model to the CO spectra. The three vertical dashed lines in each histogram indicate the 16, 50, and 84 th percentiles. }
    \label{fig:corner} 
\end{figure*}

Figure \ref{fig:corner} presents the marginal probability distribution of the spectral fitting results.
$T_0$ and intensity reduction factors are successfully constrained to be $\sim 45 $ K, and $\sim 0.8$ for both Band 6 and Band 7.
The gas surface density and CO/H$_2$ were not constrained.
This is reasonable because these quantities are completely degenerate in the observed line intensities.
%To break this degeneracy, optically thin CO isotopologue lines are needed as \citet{yosh22} did, which is out of the scope of this paper.
{ This degeneracy can be broken by using an optically thin CO isotopologue line that constrains the CO column density \citep[$\propto \Sigma_{g,0} X_{\rm CO} $, ][]{yosh22}.
However, since we aim to constrain the intensity reduction factors here and they are independent of the gas surface density and CO/H$_2$ (Figure \ref{fig:corner}), we do not include an isotopologue line in this analysis.
}
Synthetic spectra using randomly sampled parameters from the chain are also plotted in Figure \ref{fig:spec}. 
Both of the CO spectra are reproduced well by pressure broadening, strengthening the evidence that pressure broadening actually appears significantly in the inner region of the TW Hya disk \citep{yosh22}.

\begin{figure}[hbtp]
    \epsscale{1.0}
    \plotone{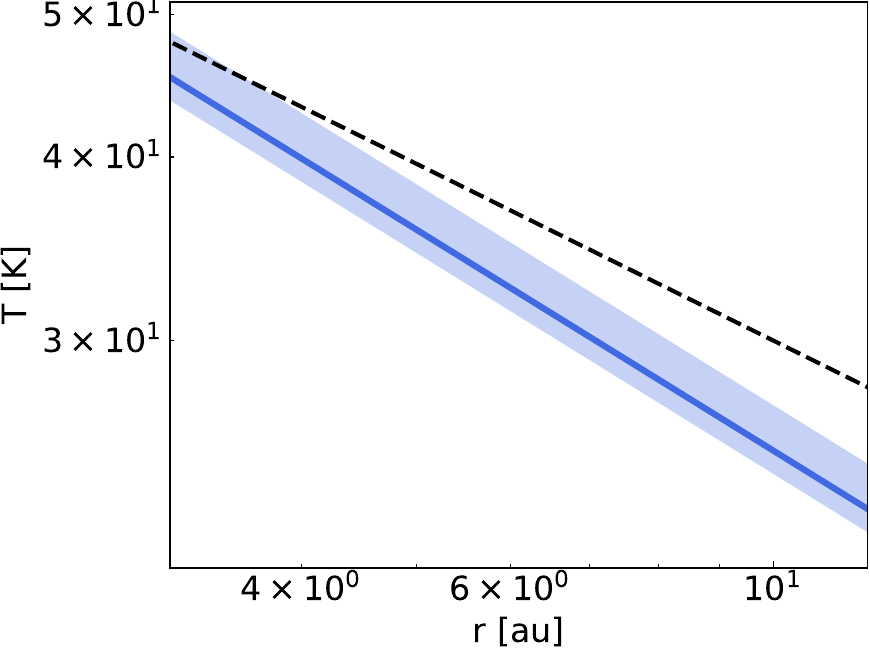}
    \caption{ Estimated temperature profile (Blue). The black dashed line is the best-fit temperature profile by \citet{ueda20} for comparison. }
    \label{fig:temp} 
\end{figure}

To evaluate the uncertainty due to the absolute flux error, we repeated this MCMC fitting by re-scaling the intensity within the $1\sigma$ flux accuracy (see Appendix \ref{sec:fluxscaling}).
In practice, the CO J=2-1 spectrum was re-scaled by $1\pm 0.027$ without re-scaling the CO J=3-2 spectrum. Then, similarly, the CO J=3-2 spectrum was re-scaled by $1\pm 0.025$ without re-scaling the CO J=2-1 spectrum.
We determined the maximum and minimum values of the temperature $T_0$ from these four additional trials.
As a result, $T_0$ with uncertainties is estimated to be $45 ^{+3}_{-2}$ K.
In Figure \ref{fig:temp}, we plot the resultant radial temperature profile compared with the previous estimate by \citet{ueda20}.
Our results are slightly smaller than the best-fit profile of \citet{ueda20}.
This is likely because \citet{ueda20} used the DSHARP dust opacity model \citep{birn18} which has a higher albedo and can reduce the emission more.
The different power law index between this study and \citet{ueda20} may also change the best-fit $T_0$ value.
However, it does not significantly affect the resultant intensity reduction factors as long as the total continuum flux is conserved.
Indeed, we re-ran the fitting with a power law index of 0.4, which is the same as \citet{ueda20}, and obtained the same intensity reduction factors within the uncertainty range as the first case.
%since we always model the whole disk within the beam rather than a specific point

\begin{figure}[hbtp]
    \epsscale{1.0}
    \plotone{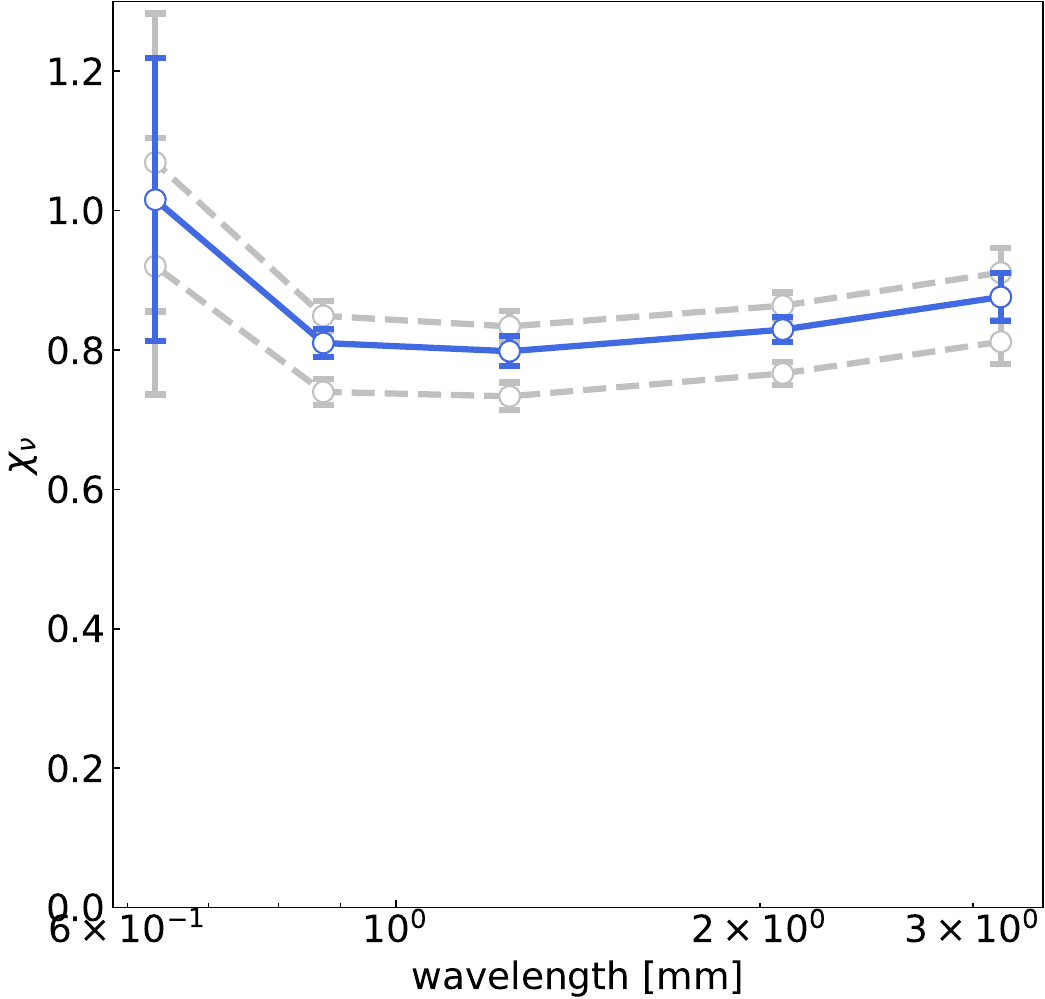}
    \caption{ { Intensity reduction factors at Band 3 ($\lambda = 3.2\ {\rm mm}$), 4 ($2.1\ {\rm mm}$), 6 ($1.2\ {\rm mm}$), 7 ($0.87\ {\rm mm}$), and 8 ($0.63\ {\rm mm}$) at the inner region ($r<6\ {\rm au}$) of the TW Hya disk.} The grey points and dashed lines are derived by taking into account the uncertainty in the temperature.}
    \label{fig:chi_obs} 
\end{figure}

The resultant temperature profile is then inputted to the continuum SED model by setting the line optical depth to zero.
Then, we divided the SED model by the observed SED (see Figure \ref{fig:sed}) and derived the intensity reduction factors as shown in Figure \ref{fig:chi_obs}.
The grey points and dashed lines correspond to the intensity reduction factors derived by using the upper- and lower-limit of the temperature profile.
The intensity-reduced factors take values of $\sim 0.7-1.0$.

\begin{figure}[hbtp]
    \epsscale{1.0}
    \plotone{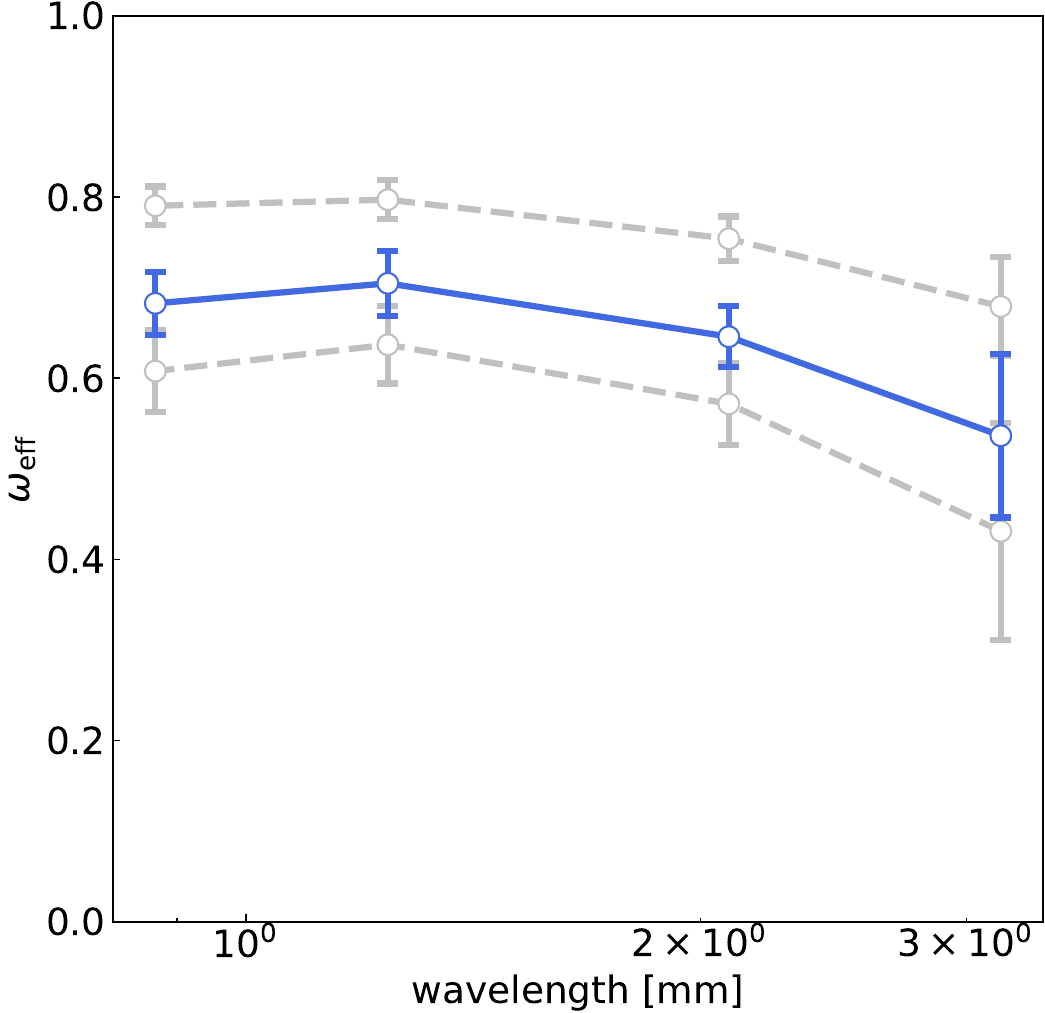}
    \caption{ Albedo spectrum of the inner region ($r<6\ {\rm au}$) of the TW Hya disk. The grey points and dashed lines are derived by taking into account the uncertainty in the temperature. }
    \label{fig:albedo} 
\end{figure}

Finally, this spectrum of $\chi_\nu$ can be converted to the effective albedo via the $\chi-\omega$ relation obtained by the RADMC-3D simulation (Figure \ref{fig:chiomega}).
At Band 8, the intensity-reduced factor is close to $\sim1.0$ with large uncertainty, which prevents from determining the albedo.
Therefore, we eliminated Band 8 in the following analysis.
The resultant dust albedo spectrum at the inner region ($r<6\ {\rm au}$) of the TW Hya disk is shown in Figure \ref{fig:albedo}.
The error bars of each line are estimated by propagating the error bars of $\chi_\nu$.
These results straightforwardly show, without assuming any dust opacity model, that the albedo in the TW Hya disk is high and the scattering-induced intensity reduction indeed works \citep{miya93, zhu19, ueda20}.
This implies that, as \citet{zhu19} indicated, the optically thin assumption of dust emission could be actually failed in some disks, and therefore, their dust mass would be underestimated.

\section{Constraining the Dust Property} \label{sec:dustmodels}

In this section, we aim to constrain the dust property from the observed albedo spectrum.
First, we show a comparison with some dust models in the literature. Second, we aim to fine-tune the DSHARP dust model to fit the observations. Finally, we run grid models with relatively large number of freedom to check which parameters are well-constrained.
In the following, we compared the observed and modeled albedo spectrum in terms of $\omega_{\rm eff}$.
The effective scattering opcaity, $\kappa_s^{\rm eff}$ in Equation (\ref{eq:sca_opa}), is assumed to be $\kappa_s^{\rm eff} = (1-g) \kappa_s$, where $g$ and $\kappa_s$ are the forward-scattering parameter and scattering opacity, respectively.
This is a good approximation for optically thick media \citep{ishi78, birn18}.
To treat the full anisotropic effect, it is necessary to perform disk-specific radiative transfer calculations for each dust model, which is beyond the scope of this paper.

\subsection{Comparison with Frequently-Used Dust Models}
\begin{figure*}[hbtp]
    \epsscale{1.2}
    \plotone{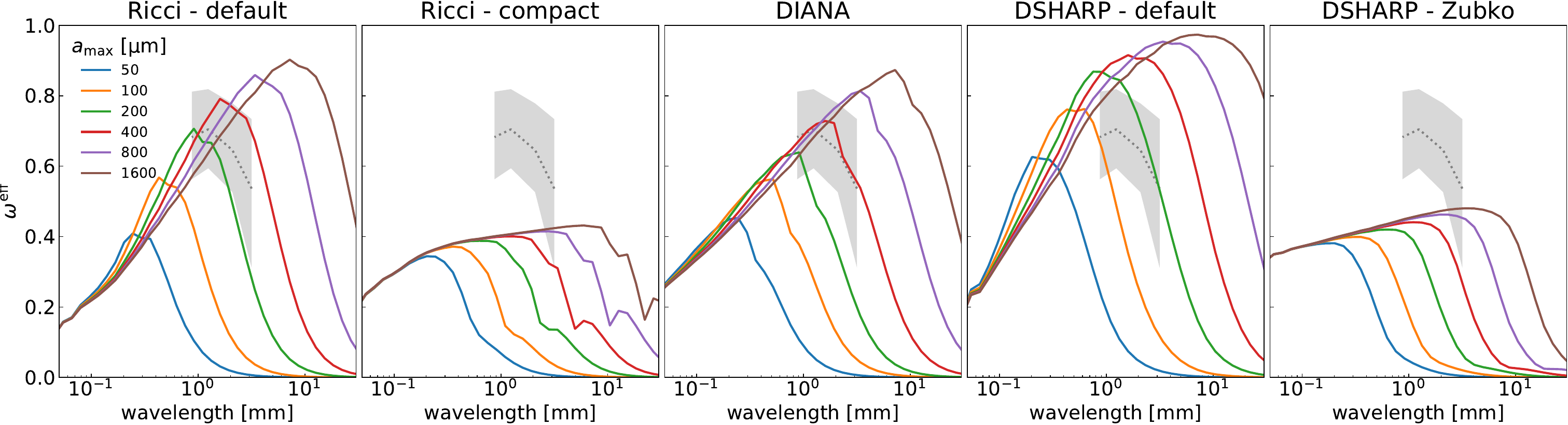}
    \caption{ Comparison of albedo spectra created from dust models on the literature for various maximum dust sizes (color lines) with ones derived from observations (grey area). }
    \label{fig:comparison} 
\end{figure*}

{ We generated effective albedo spectra for five frequently used dust opacity models; Ricci default, Ricci compact, DIANA, DSHARP default, and DSHARP Zubko.
The Ricci default model is originally proposed by \citet{ricc10}.
%In addition to the default model, we introduce the Ricci compact model, where the porosity is set to zero in the Ricci default model.
In addition, we introduce the Ricci compact model, which is similar to the Ricci default model but with zero porosity.
This is motivated by \citet{delu24}, where they found that the Ricci compact model better reproduces the spectral indices obtained by disk survey observations.
We also compare the observations with the DIANA model that is proposed by \citet{woit16}.
Additionally, we show the DSHARP default model \citep{birn18} and DSHARP Zubko model that is created by replacing the refractory organics in the default model with the amorphous carbon in \citet{zubk96}.
}
The detailed model assumptions are listed in Table \ref{tab:dustmodel}.
{ We calculated the absorption and effective scattering opacities using {\tt optool} \citep{optool} for the DIANA models, and {\tt dsharp\_opac} \citep{birn18} for the Ricci and DSHARP models, respectively.
Note that we used the Bruggeman mixing rule throughout the calculation in {\tt dsharp\_opac}  \citep{bohr98, birn18}.}
With changing the maximum dust size $a_{\rm max}$, the albedo spectra are overplotted in Figure \ref{fig:comparison}.
{ The Ricci default and DIANA model with $a_{\rm max} = 200-400\ \mu m$ can explain the observed spectrum well.
The DSHARP default model is also roughly consistent with the observed spectrum. However, the peak albedo is slightly higher than the observed level.
On the other hand, the Ricci compact model and the DSHARP Zubko model cannot reproduce the observed high albedo for any $a_{\rm max}$.}

%\begin{rotatetable}
\begin{deluxetable*}{lccccc}
    %\centering
    \caption{{ Dust models considered in this paper\label{tab:dustmodel} }}

   % \hline \hline
      \tablehead{  \colhead{Model} & \colhead{Grain geometry} & \colhead{Ice} & \colhead{Mineral} & \colhead{Carbon} & \colhead{Porosity} } %\hline
   % \hline
    \startdata
        Ricci - default \tablenotemark{a} & Sphere  & Water\tablenotemark{b}, 30\% & Astronomical silicates\tablenotemark{c}, 20\%  & Amorphous carbon\tablenotemark{d}, 10\%  & 40\% \\ \hline
        Ricci - compact \tablenotemark{a,e} & Sphere  & Water\tablenotemark{b}, 50\% & Astronomical silicates\tablenotemark{c}, 33\%  & Amorphous carbon\tablenotemark{d}, 17\%  & 0\% \\ \hline
        DIANA\tablenotemark{f} & Hollow sphere  & N/A & Pyroxyn\tablenotemark{g}, 60\%  & Amorphous carbon\tablenotemark{h}, 15\%  & 25\% \\ \hline
        DSHARP\tablenotemark{i} - default & Sphere & Water\tablenotemark{j}, 36 \%  & \shortstack{ Astronomical silicates\tablenotemark{k}, 17 \%  \\   Troilite\tablenotemark{l}, 3\%  } & Refractory organics\tablenotemark{l}, 44\%  & 0\%  \\ \hline
        DSHARP\tablenotemark{i} - Zubko & Sphere & Water\tablenotemark{j}, 36 \%  & \shortstack{ Astronomical silicates\tablenotemark{l}, 17 \% \\ Troilite\tablenotemark{l}, 3\%  } & Amorphous carbon\tablenotemark{d}, 44\%  & 0\%  \\
\enddata

    %ricci
    \tablenotetext{a}{\citet{ricc10}}
    \tablenotetext{b}{\citet{warr84}}
    \tablenotetext{c}{\citet{wein01}}
    \tablenotetext{d}{ACH-sample of \citet{zubk96} }
    \tablenotetext{e}{\citet{delu24}} 
    
    %DIANA
    \tablenotetext{f}{\citet{woit16}} 
    \tablenotetext{g}{\citet{dors95}}
    \tablenotetext{h}{BE-sample of \citet{zubk96} }

    %DSHARP
    
    \tablenotetext{i}{\citet{birn18}}
    \tablenotetext{j}{\citet{warr08}}
    \tablenotetext{k}{\citet{drai03}}
    \tablenotetext{l}{\citet{henn96}}

\end{deluxetable*}

\subsection{Fine-Tuning the DSHARP model} \label{sec:finetune}
\begin{figure}[hbtp]
    \epsscale{1.0}
    \plotone{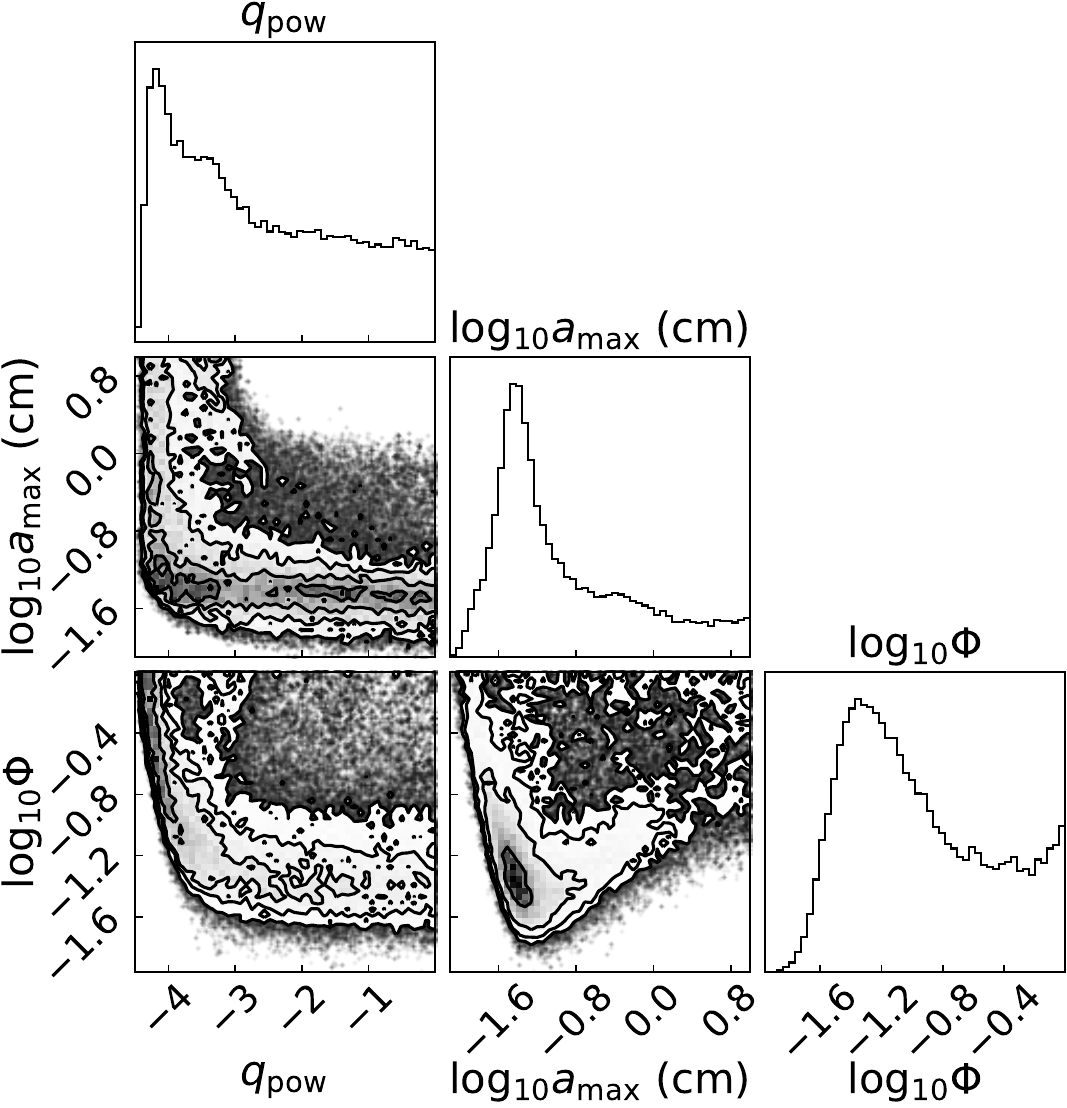}
    \caption{ Corner plot of the MCMC fitting of dust model to the derived dust albedo spectrum. }
    \label{fig:dustmodelfitting} 
\end{figure}

Although both the DIANA and DSHARP default models are consistent with the observed albedo spectrum, we try to fine-tune the DSHARP model to match the observation better.
The DSHARP default model assumes the power law index of the grain size distribution $q_{\rm pow} = -3.5$ and porosity $p = 0$.
However, in reality, $q_{\rm pow}$ and $p$ (or volume filling factor $\Phi \equiv 1-p$) can be different from those values.
Therefore, with these parameters as well as $a_{\rm max}$ being free parameters, we generated albedo spectra using {\tt dsharp\_opac} \citep{birn18} and fitted them to the observed result with {\tt emcee} \citep{emcee}.
Here, we put the upper and lower limit of the albedo spectrum as the uncertainty for fitting as shown in the grey shaded region of Figure \ref{fig:comparison}.
The resultant marginal distribution is plotted in Figure \ref{fig:dustmodelfitting}.
$a_{\rm max}$ would be constrained to be $330^{+350}_{-120}\ {\rm \mu m}$.
Since the peak of the marginal distribution is not very clear, the value with uncertainty is determined by the peak and half of the peak in the distribution.
The marginal distribution for $q_{\rm pow}$ exhibits only a weak peak around $q_{\rm pow} = -4.2$ with a heavy tail. 
We can only give a rough lower limit of $\Phi$ of 2\%, or upper limit of $p < 98 \%$ at the primary half of the highest peak of the marginal distribution.

It would be worthwhile to see dependencies of albedo spectra on these parameters.
\begin{figure*}[hbtp]
    \epsscale{1.1}
    \plotone{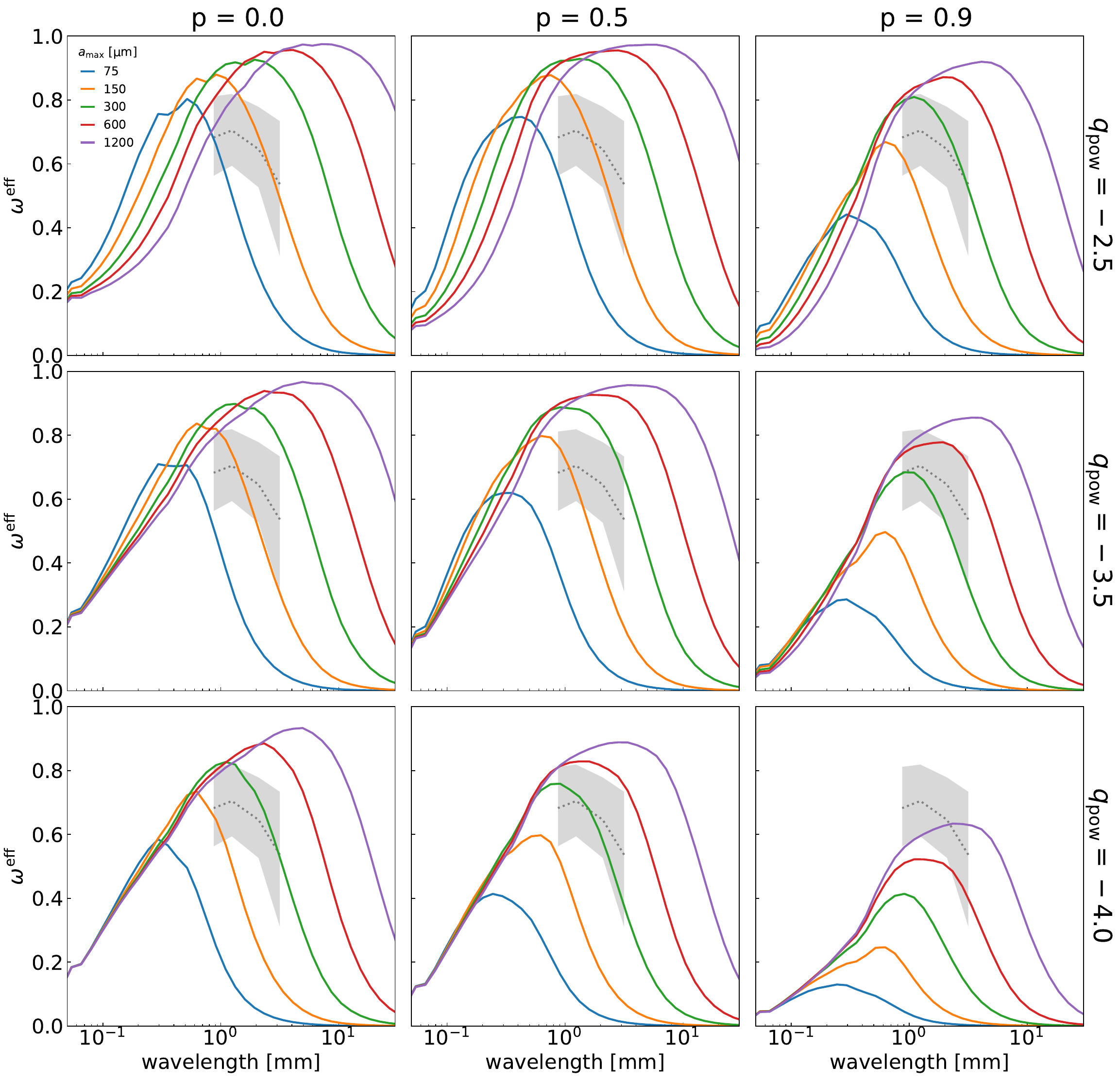}
    \caption{ Dependencies of the albedo spectrum on $a_{\rm max}$, $q_{\rm pow}$ and $p$ in the case of the DSHARP default composition model (color lines). Grey area shows constraints from observations. }
    \label{fig:thiswork_ds} 
\end{figure*}
In Figure \ref{fig:thiswork_ds}, we show albedo spectra for various $a_{\rm max}$, $q_{\rm pow}$ and $p$.
As expected, the peak wavelength of the albedo spectrum is sensitive to $a_{\rm max}$ \citep[e.g.,][]{ueda20}. 
A steeper power law index of the grain size distribution would lower the albedo.
This is because larger grains whose albedo is low dominate the total opacities.
In addition, porous dust with a steep power law index of the grain size distribution cannot explain the high albedo \citep[e.g.,][]{taza19}.
%If we adopt the DSHARP default composition, the models with ($a_{\rm max}$, $q_{\rm pow}$, $p$) = ( $300\ \mu m$, $0.5$, $-4.0$ ), ( $300\ \mu m$, $0.9$, $-2.5$ ), or ( $300\ \mu m$, $0.9$, $-3.5$ ) are candidates for the observed dust in this model grid.

\subsection{Composition-independent constraints on $q_{\rm pow}$, $a_{\rm max}$, and $p$}
\label{sec:compoind}
\begin{figure*}[hbtp]
    \epsscale{1.0}
    \plotone{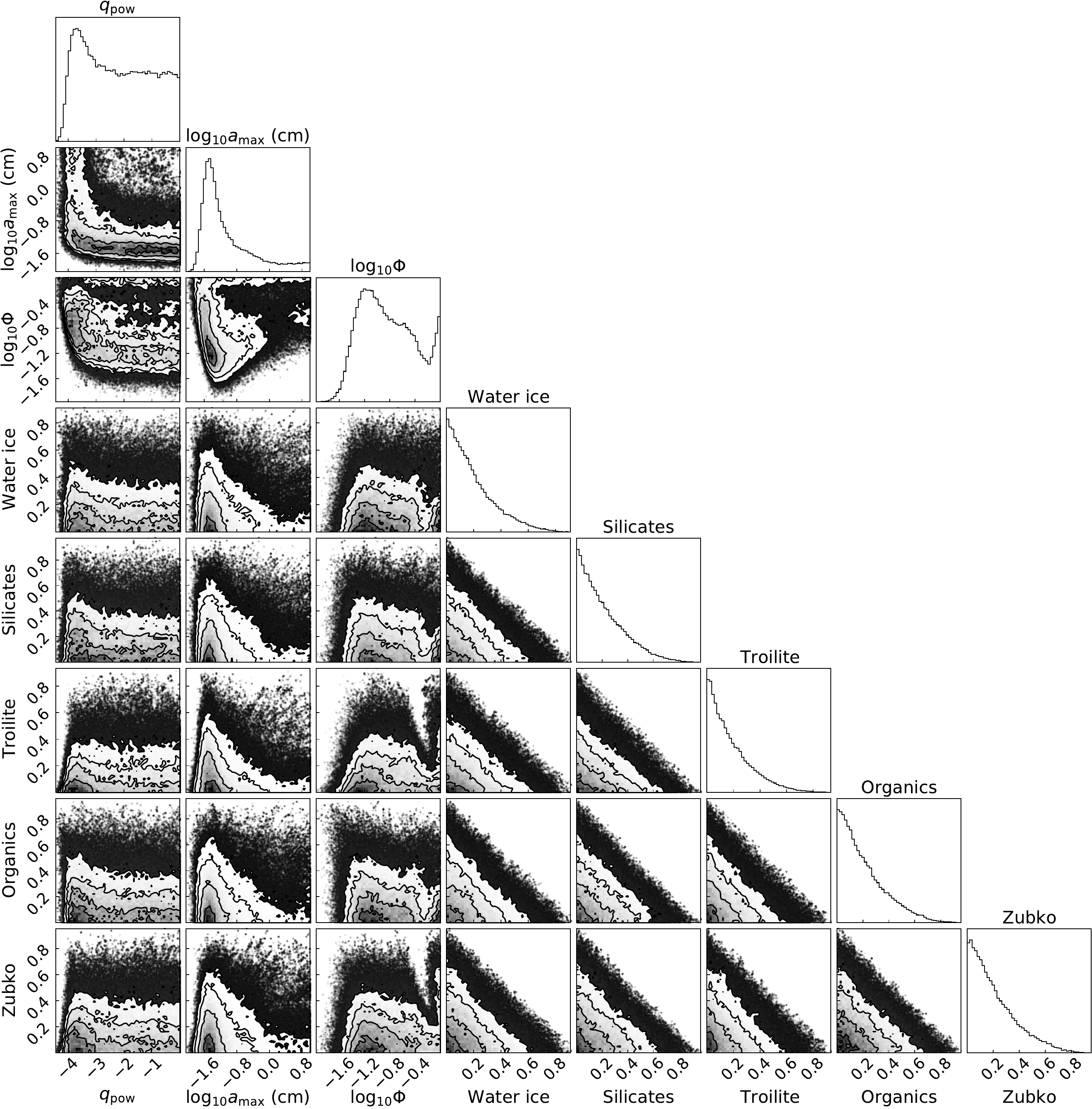}
    \caption{ Corner plot of the MCMC fitting of dust model to the derived dust albedo spectrum. In this fitting, we kept the composition parameters free. }
    \label{fig:dust_corner_all} 
\end{figure*}

The above results imply that the dust grains should not be highly porous ($p \lesssim 98\%$) and their size distribution should not be very steep ($q_{\rm pow} \gtrsim -4.2$).
To investigate whether our fitting results depend on the composition, we performed another MCMC sampling with additional parameters on volume fractions of material; water ice, astronomical sillicates, troilite, refractory organics, and Zubko carbon (Figure \ref{fig:dust_corner_all}).
As a result, none of the composition parameters are well-constrained with a rough upper limit of $\sim50\%$; any material cannot exceed $\sim50\%$ of the total volume.
Nevertheless, the size distribution and porosity, $a_{\rm max}$, $q_{\rm pow}$, $p$, show similar marginal distribution to the case when the composition is fixed.
In the same way as Section \ref{sec:finetune}, we calculated the representative parameters to be ($a_{\rm max}$, $q_{\rm pow}$, $p$) = ( $340^{+180}_{-120}\ \mu m$, $>-4.1$, $<0.96$ ).

\section{Discussion}  \label{sec:disc}

\subsection{Self-consistency of the analysis} \label{sec:self_consistency}
\begin{figure}[hbtp]
    \epsscale{1}
    \plotone{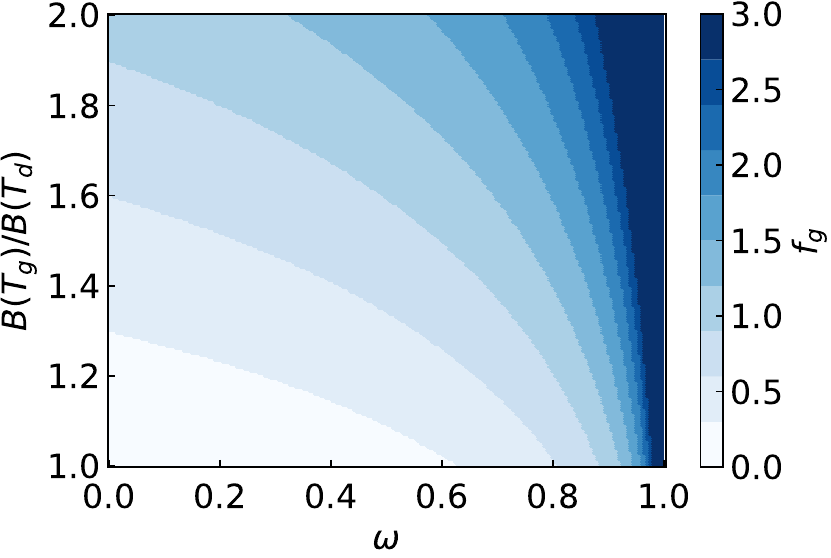}
    \caption{ Line emerging factor $f_g$ as a function of $\omega$ and $B(T_g)/B(T_d)$.}
    \label{fig:fg} 
\end{figure}

First, we check the consistency of the results with an assumption of $B(T_g)/B(T_d)=1$ (Section \ref{sec:2.2}). 
We plot the line emerging factor $f_g$ as a function of $\omega$ and $B(T_g)/B(T_d)$ in Figure \ref{fig:fg}.
This figure shows that we can observe optically thin line emission above optically thick continuum emission only when the albedo is high and/or the vertical temperature gradient is large.
%For example, if $\omega=0.8$ and $B(T_g)/B(T_d) = 1$, the line emerging factor $f_g$ is $\sim 0.6$.
%To maintain the same intensity for fixed line optical depth without scattering,  $B(T_g)/B(T_d) \sim 1.6$ is needed.
%Figure \ref{fig:fg2} illustrates $f_g$ as a function of $\omega$ when $T_g = T_d$.
%In other words, if we can constrain $\tau_g$ independently of the continuum emission, the albedo can be determined from observations.

The CO spectral fitting suggests the intensity reduction factors of $\sim0.8$, or albedos of $\sim0.7$.
According to Figure \ref{fig:fg}, the temperature difference between the dust and gas layer of $\sim1.4$ is required to explain the spectra with zero-albedo for the same optical depths.
In protoplanetary disks, however, the temperature is expected to be almost constant along the vertical axis within one or two gas scale heights from the midplane \citep{dull02, pint09, inou09}.
Additionally, the dust and gas are thermally coupled when the density is sufficiently high as in the region we are tracing.
%Therefore, under the assumption that a molecular line originates near the midplane, $f_g$ only depends on the albedo $\omega$. 
%However, it is unlikely that the vertical temperature structure has such a steep gradient near the midplane \citep{dull02, inou09}.
Therefore, the high albedo scenario is more likely to explain the observations rather than a steep temperature gradient.

We also assumed the layered dust and gas structure.
If the dust grains are well mixed with gas and the dust emission is optically thick, the line emission cannot be observed unless the line emission is optically thick and there is a steep temperature gradient \citep{bosm21}.
If the line wings were optically thick in reality, the intensity would only reflect the local temperature at each height that is determined by the optical depth as a function of velocity shift from the line center, which requires a specific temperature structure with a steep gradient.
However, both the observed CO spectra are well fitted by the optically thin line wings with a constant temperature (Figure \ref{fig:spec}), suggesting the optically thin assumption is valid.

To make the dust grains and gas layered, vertical settling of dust grains is needed.
Assuming that the vertical settling and turbulent diffusion are balanced, the fraction of the dust scale height to the gas scale height $f_H$ is expressed as 
\begin{equation}
    f_H = \left(  1 + \frac{{\rm St}}{\alpha} \frac{1+2{\rm St}}{1+{\rm St}} \right)^{-0.5},
\end{equation}
where ${\rm St}$ and $\alpha$ are the Stokes number and turbulence parameter, respectively \citep[e.g.,][]{shak73, dubr95, youd07, okuz12}.
The Stokes number is defined as
\begin{equation}
\label{eq:stokes}
    {\rm St} = \frac{\pi \rho a}{2\Sigma_g},
\end{equation}
where $\rho$ is the dust material density, and $a$ is the dust radius.
{ If we substitute $\rho=2 \ {\rm g\ cm^{-3}}$ \citep{woit16, birn18}, $a=340\rm {\rm \mu m}$, and $\Sigma = 1200\ {\rm g\ cm^{-2}}$ at $r=5$ au \citep{yosh22}, ${\rm St}$ can be calculated to be $\sim 9\times 10^{-5}$.}
The turbulence parameter $\alpha$ was estimated to be $\sim 10^{-4}$ by \citet{yosh22}, resulting in $f_H \sim 0.7$. Note that isotropic turbulence is assumed here.
This value is not very small but still consistent with the settling scenario.
At the height of $0.7\times$ gas scale height, the ${\rm H_2}$ gas volume density is $e^{-0.5 \times (0.7)^2} \sim 0.8 $ of the value at the midplane, which does not prevent observable pressure-broadened line wings since the pressure-broadened line wings are sensitive to the ${\rm H_2}$ gas volume density.

\subsection{Constraints on dust properties}
Although it is implied that both of the DIANA and DSHARP default models can reproduce the observed albedo spectrum, it is difficult to determine the dust composition directly.
However, if the refractory organics were replaced by the amorphous carbon in the DSHARP default model, the high albedo cannot be explained.
This is because the amorphous carbon proposed by \citet{zubk96} has a relatively low albedo compared to other materials (Kataoka et al. in prep).
Troilite has similar optical constants, and therefore, it should not be a major material of dust grains.
%The rough upper limit of these materials can be given as $\sim50\%$ (see Section. \ref{sec:compoind}).

Even after freeing all dust property parameters, the maximum grain size $a_{\rm max}$ of $\sim 340\ {\rm \mu m}$ is obtained by the fitting, and this is still comparable with the previous analysis by \citet{ueda20} and \citet{maci21}, where fixed dust composition models are used.
The derived dust size may suggest efficient fragmentation of dust particles.
With the latest physical parameters of the TW Hya disk, we revisit the discussion about the fragmentation of dust particles in \citet{ueda20}.
Following their Equation (7), if the dust size is determined by the turbulence-induced collisional fragmentation, the fragmentation velocity $v_{\rm frag}$ can be expressed as
\begin{equation}
    v_{\rm frag} = \sqrt{ 3 {\rm St_{max}} \alpha } c_s,
\end{equation}
where ${\rm St_{max}}$ is the Stokes number for the maximum grain size.
$c_s$ is the sound speed and estimated to be $0.4\ \kms$ at the temperature of $64$ K at $r=5$ au and the mean molecular weight of 2.37.
Substituting ${\rm St_{max}} = 1.8 \times 10^{-4}$ (Equation \ref{eq:stokes}), $\alpha = 10^{-4}$, and $c_s = 0.4\ \kms$, we obtain $v_{\rm frag} \sim 0.08\ {\rm m\ s^{-1}}$.
This low $v_{\rm frag}$ suggests that the dust grains are fragile.
For example, this is consistent with the value experimentally measured for the CO$_2$-H$_2$O ice mixture \citep{musi16, okuz19}.
Indeed, it is suggested that the CO gas in the TW Hya disk is significantly depleted \citep[e.g.,][]{berg13, bosm19, zhan19, yosh22}, and a part of the depleted CO would be converted to CO$_2$ ice on the dust grains \citep{krij20}.
However, it would be notable that $v_{\rm frag}$ depends on other parameters such as the monomer size \citep{arak21}.
{ As an alternative idea to the fragmentation-limited dust growth, such a small maximum grain size might suggest that the dust size is actually limited by the bouncing barrier \citep{domi24}.}

$q_{\rm pow}$ is in line with \citet{math77} who found $q_{\rm pow} = -3.5$ in the interstellar medium.
This is also consistent with the measurements in the HD 163296 disk \citep{doi23}.
Porosity $p$ is consistent with the ballistic particle-cluster aggregation \citep[BPCA;][]{muka92, koza92, shen08}.

In our analysis, the uncertainty of the albedo spectrum is dominated by the absolute flux uncertainty.
If future observations can measure the absolute flux more robustly, it may become possible to more accurately constrain the dust property.
For example, three models in Figure \ref{fig:comparison} imply that the composition may affect the slope of the albedo spectrum in relatively shorter wavelengths.
Far-infrared or sub-millimeter observations with a high flux accuracy would be a key to constraining the dust property in protoplanetary disks using albedo spectra.

\subsection{Diversity of Dust Properties}
{
In this study, we focused on a limited region of a single disk, therefore, it is challenging to generalize the results.
However, while it is natural that the maximum dust size and porosity vary as a function of the radial location, for instance, the fragmentation velocity and composition might be somewhat universal, especially in the cold outer regions, since the dust grains originated from the interstellar medium and experienced similar physical conditions.

In this context, it is interesting to compare our results with \citet{delu24}.
\citet{delu24} found that the Ricci compact model matches observations better than the DSHARP default model, while we found the opposite.
This tension could be just due to different quantities that the two studies focus on; we discuss the albedo but \citet{delu24} may have a sensitivity on the total (absorption plus scattering) opacity since they study the total flux which is presumably dominated by optically thin emission.
Therefore, a composition with a high absorption opacity as in the Ricci compact model, and a high scattering opacity that can maintain the high albedo could be consistent with both results, although it is unclear that such material actually exists.
A more attractive scenario is that there is a diversity in dust composition (or porosity) among disks or regions in a disk. At least, the dust grains in the inner region of the TW Hya disk may be different from the average dust properties in disks that \citet{delu24} studied. This could be due to chemical and physical processes in the disk considering that TW Hya is relatively old \citep[$3-10$ Myr; ][]{barr06, vacc11}.
}

\section{Conclusion}  \label{sec:conc}

It is crucial to understand dust grains in protoplanetary disks for revealing planet formation processes.
In this study, we measured the wavelength-dependent dust scattering albedo, or albedo spectrum, using millimeter dust continuum observations.
If the albedo is high, it is known that the scattering reduces the intensity of the dust continuum emission.
The albedo and dust temperature are { typically} degenerate, which prevents measuring both of them without assuming a dust model.
Our idea to break this degeneracy is to use two or more transition lines from the same molecules as a thermometer of the disk midplane.
We applied the method to the pressure-broadened line wings of CO $J=2-1$ and $J=3-2$ in the inner region of the TW Hya disk as the broad line wings exceptionally bring a high signal-to-noise ratio.
We constructed an emission model, fitted it to the spectra, and determined the temperature at the inner $r < 6$ au region.
Then, with multi-band continuum observations spanning from 0.9 mm to 3 mm, we directly constrained the albedo spectrum for the first time.
The albedo is high and ranges $\sim0.5-0.8$.
Additionally, it is proved that the scattering-induced intensity reduction indeed works in the inner region of the TW Hya disk.

{ Among some previously proposed dust models, the Ricci default, DIANA, and DSHARP default models are broadly consistent with the observed albedo spectrum, although the Ricci compact and DSHARP Zubko models are inconsistent.}
We then search for the dust size distribution, porosity, and composition that can reproduce the albedo spectrum.
While the composition cannot be constrained, the maximum dust size, power law index of the dust size distribution, and porosity are constrained to be $340^{+180}_{-120}\ \mu m$, $>-4.1$, and $<0.96$, respectively.
These results are in line with an efficient fragmentation of dust particles with a fragmentation velocity of $\sim 0.08\ {\rm m\ s^{-1}}$.

{ In this paper, we focused on an inner region of a single protoplanetary disk.
Therefore, it is not clear how much the results here are universal. 
Indeed, the Ricci compact model is excluded in our analysis while it matches the survey observations well \citep{delu24}. Although their study is sensitive to different quantities from our study, this tension might imply a diversity of dust properties among regions and/or disks.
We note that the method proposed in this study is in principle applicable to other disks, which enables us to obtain a comprehensive view of the albedo and dust properties.
}

Before the detailed analysis, we carefully checked the uncertainty of the flux scaling of the images.
Our results suggest that the absolute flux uncertainty on the image plane can be generally much larger than the nominal absolute flux uncertainty of the calibrators in the ALMA technical handbook.
Future observations with high flux accuracy, which can be achieved by using four or more flux calibrators \citep{fran20}, would be significantly important to understand the dust property.

\begin{acknowledgments}
This paper makes use of the following ALMA data: 2015.1.00686.S, 2016.1.00629.S, 2018.1.00980.S, 2018.A.00021.S, 2015.1.01137.S, 2016.1.01399.S, 2016.1.00229.S, 2018.1.01218.S, 2015.A.00005.S, 2015.1.00845.S, 2016.1.00842.S, 2016.1.00440.S, 2012.1.01012.S,	2017.1.01199.S,	2017.1.01587.S,	2019.1.00769.S, 	2016.1.00440.S,	2018.1.00810.S,	2019.1.01153.S, 2013.1.00387.S,	2016.1.00311.S,	2016.1.01046.S,	2016.1.01375.S	2018.1.00167.S,	2018.A.00021.S,	2019.1.01177.S, 2012.1.00400.S,	2013.1.00192.S,	2013.1.00198.S,	2013.1.01397.S,	2016.1.00311.S,	2016.1.00629.S,	2018.1.00460.S,
2012.1.00422.S,	2013.1.00196.S,	2013.1.00902.S,	2015.1.00308.S,	2016.1.00464.S,	2016.1.01495.S,	2018.1.00980.S.
This work was supported by Grant-in-Aid for JSPS Fellows, JP23KJ1008 (T.C.Y.).
\end{acknowledgments}

%% To help institutions obtain information on the effectiveness of their 
%% telescopes the AAS Journals has created a group of keywords for telescope 
%% facilities.
%
%% Following the acknowledgments section, use the following syntax and the
%% \facility{} or \facilities{} macros to list the keywords of facilities used 
%% in the research for the paper.  Each keyword is check against the master 
%% list during copy editing.  Individual instruments can be provided in 
%% parentheses, after the keyword, but they are not verified.

\vspace{5mm}
\facilities{ALMA}

%% Similar to \facility{}, there is the optional \software command to allow 
%% authors a place to specify which programs were used during the creation of 
%% the manuscript. Authors should list each code and include either a
%% citation or url to the code inside ()s when available.

\software{astropy \citep{astropy}, CASA \citep{mcmu07} }

%% Appendix material should be preceded with a single \appendix command.
%% There should be a \section command for each appendix. Mark appendix
%% subsections with the same markup you use in the main body of the paper.

%% Each Appendix (indicated with \section) will be lettered A, B, C, etc.
%% The equation counter will reset when it encounters the \appendix
%% command and will number appendix equations (A1), (A2), etc. The
%% Figure and Table counter will not reset.

\appendix

\section{Flux Scaling Using Archival Images} \label{sec:fluxscaling}
ALMA uses quasars or solar system objects to set the absolute flux scaling.
The absolute flux of such flux calibrators is estimated by observations with other instruments or some models.
According to the ALMA Cycle 10 technical handbook, the $1 \sigma$ uncertainty of the absolute flux scaling is claimed as $\sim2.5\ \%$ for Bands 3,4, and $\sim 5$ \% for Bands 6 and 7, respectively.
However, assessment of the flux accuracy using stable young stellar objects suggested that the accuracy could be poorer than $\sigma \sim 10$ \% at Band 7 \citep{fran20}.

Since our study is sensitive to the absolute flux, we checked the flux accuracy before the detailed analysis.
Many independent observations of the TW Hya disk are available on the ALMA science archive thanks to its popularity.
Therefore, we can use the TW Hya disk itself as a flux calibrator assuming that the mm-continuum emission is constant during this $\sim10$ years.
We downloaded all the available product Band 3, 4, 6 and 7 images on the ALMA science archive but excluded ones with a small maximum recoverable scale of $<3$ arcsec.
The project IDs are written in the acknowledgment section. 
Then, we re-convolved the images to match the synthesized beam to the largest one for each subset of the bands and measured the total flux with an aperture of 3 arcsec in radius from the source center specified by 2D Gaussian fitting.
Note that the synthesized beam is always smaller than the aperture.

\begin{figure}[hbtp]
    \epsscale{0.6}
    \plotone{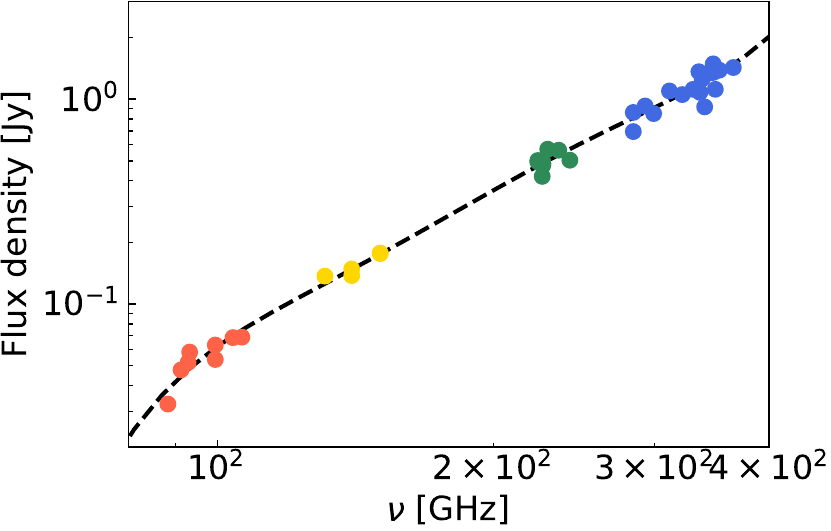}
    \caption{ Total flux measured on product images with respect to frequency. Different colors indicates different ALMA Bands. }
    \label{fig:total_sed} 
\end{figure}

\begin{figure}[hbtp]
    \epsscale{1.0}
    \plotone{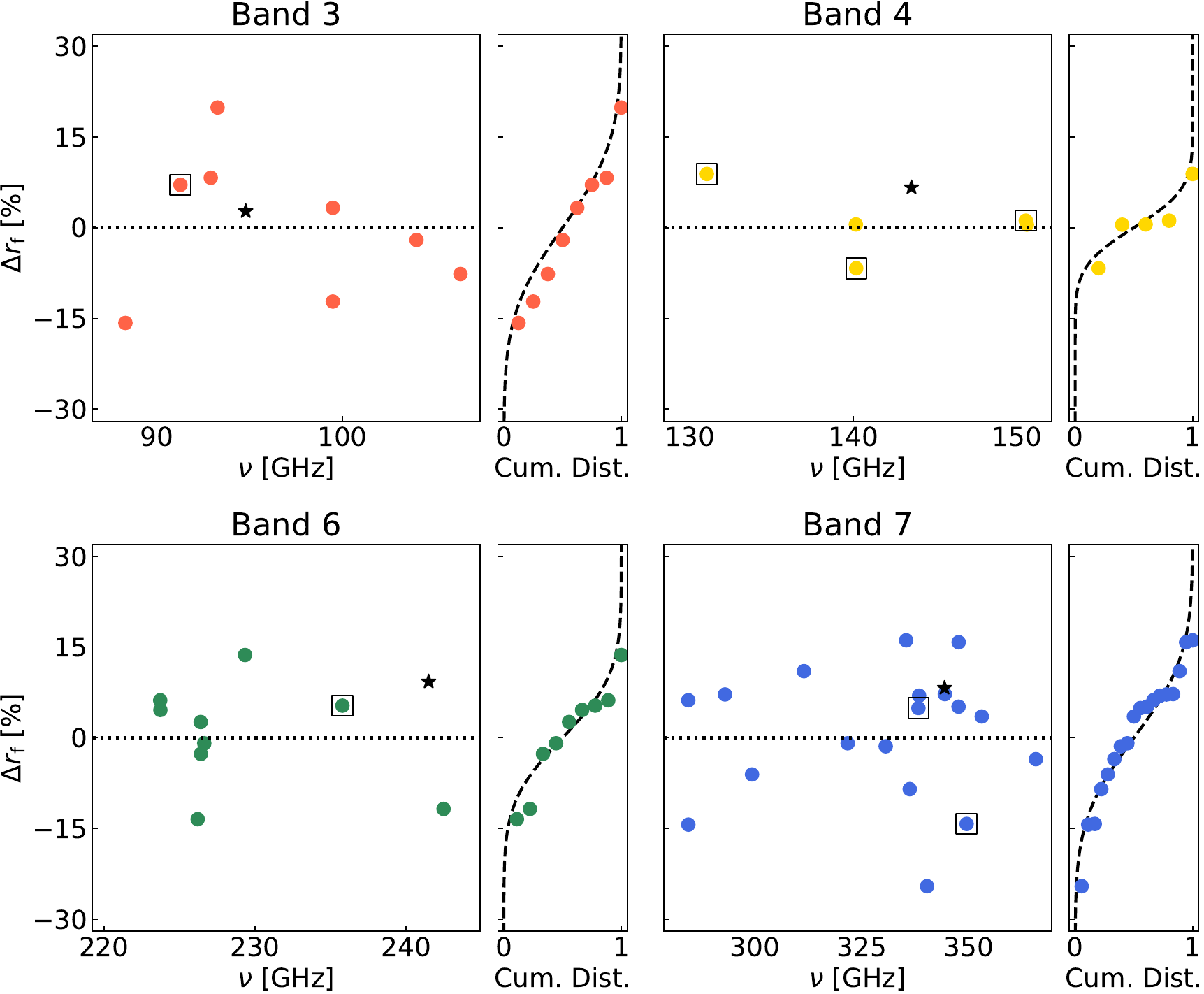}
    \caption{ Deviation of measured fluxes against fitted flux densities. Star marks measured total flux on our self-calibrated images used in the analysis in the main sections. Squares indicates the product images which correspond to the data used for the self-calibrated images. The right panels show the cumulative distribution of measured fluxes (points) and the error function with the standard deviation of the measured fluxes (dashed lines).}
    \label{fig:delta_r_f} 
\end{figure}

Figure \ref{fig:total_sed} indicates the total flux versus frequency for all bands.
We fitted these points by the fifth-degree polynomial to estimate the mean values at each frequency, which is also plotted in Figure \ref{fig:total_sed}.
To investigate the deviation of each point from the fitted flux densities, we define
\begin{equation}
    \Delta r_f(\nu) \equiv \left( \frac{F_{\rm obs}(\nu)}{ F_{\rm fit}(\nu) } - 1 \right) \times 100,
\end{equation}
where $F_{\rm obs}(\nu)$ and $F_{\rm fit}(\nu)$ are the measured flux and fitted flux density at a frequency $\nu$, respectively.
$\Delta r_f(\nu)$ for each band is plotted in Figure \ref{fig:delta_r_f}.
The standard deviation of $\Delta r_f(\nu)$ is calculated to be $\sigma \simeq 11$\%, $4.9\%$, $8.2\%$, and $11\%$ for Band 3, 4, 6, and 7, respectively.
This is generally larger than the typically adopted values (e.g., $\sigma=5\%$ for Band 6 and 7), which may call attention to users.
We also plotted the normalized cumulative distribution of $\Delta r_f(\nu)$ with the normalized error function corresponding to the calculated standard deviation.
It is clear that the distribution of $\Delta r_f(\nu)$ is Gaussian-like.
Assuming that all the product images are statistically independent of each other, we can find that the flux densities at the analyzed frequency with the confidence intervals as shown in Table \ref{tab:meanf}.
\begin{table}[hbtp]
  \caption{Derived fluxes at each frequency}
  \label{tab:meanf}
  \centering
  \begin{tabular}{ccccc}
    \hline
    Band & Frequency [GHz] & Flux density w/o selfcal [Jy] & Flux density w/ selfcal [Jy] & Uncertainty [$\%$] \\
    \hline \hline
    3 & 95  & 0.0517  & 0.0496 & 3.9 \\
    4 & 144  & 0.156 & 0.165 & 2.2 \\
    6 & 241  & 0.566 & 0.587 & 2.7 \\
    7 & 344  & 1.25  & 1.29 & 2.5 \\
    \hline
  \end{tabular}
\end{table}

In addition to the uncertainty in the flux scaling factor, the source flux might be lost due to decorrelation of phases \citep{brog18, fran20} and further altered during imaging processes.
\citet{fran20} suggested that the loss of flux can be fixed by phase self-calibration.
We overplotted the measured total flux on our self-calibrated images with stars in Figure \ref{fig:delta_r_f}.
The product images that correspond to our images are also marked with squares.
Note that our images contain other data with longer baselines, and therefore, the frequencies between the product and our images do not match necessarily.
Assuming that the flux of each data set is generally determined by the shorter baseline observations, we can correct the difference between the product and our images; the ratio of an averaged deviation of the star points to the squared points from the mean value is multiplied to the mean flux at each frequency.
The corrected flux is also plotted in the fourth column of Table \ref{tab:meanf}.
Finally, we scaled our self-calibrated high-resolution images by comparing the total flux of our images and the mean value.

For the line image cubes of CO $J=2-1$ and $3-2$, there may be additional uncertainties due to insufficient CLEAN, which is also known as the \citet[][JvM]{jors95} effect.
Although we applied the JvM correction \citep{jors95, czek21}, it is suggested that the correction factor cannot be defined in practice \citep{casa22}.
Therefore, we rescaled the resultant spectra in the same way as the continuum images.
The continuum levels of the line spectra are determined by fitting the emission model (Section \ref{sec:emimod}) to the spectra without re-scaling.

\bibliography{sample631}{}

\begin{thebibliography}{}
\expandafter\ifx\csname natexlab\endcsname\relax\def\natexlab#1{#1}\fi
\providecommand{\url}[1]{\href{#1}{#1}}
\providecommand{\dodoi}[1]{doi:~\href{http://doi.org/#1}{\nolinkurl{#1}}}
\providecommand{\doeprint}[1]{\href{http://ascl.net/#1}{\nolinkurl{http://ascl.net/#1}}}
\providecommand{\doarXiv}[1]{\href{https://arxiv.org/abs/#1}{\nolinkurl{https://arxiv.org/abs/#1}}}

\bibitem[{{Andrews} {et~al.}(2016){Andrews}, {Wilner}, {Zhu}, {Birnstiel}, {Carpenter}, {P{\'e}rez}, {Bai}, {{\"O}berg}, {Hughes}, {Isella}, \& {Ricci}}]{andr16}
{Andrews}, S.~M., {Wilner}, D.~J., {Zhu}, Z., {et~al.} 2016, \apjl, 820, L40, \dodoi{10.3847/2041-8205/820/2/L40}

\bibitem[{{Arakawa} \& {Krijt}(2021)}]{arak21}
{Arakawa}, S., \& {Krijt}, S. 2021, \apj, 910, 130, \dodoi{10.3847/1538-4357/abe61d}

\bibitem[{{Armitage}(2010)}]{armi10}
{Armitage}, P.~J. 2010, {Astrophysics of Planet Formation}

\bibitem[{{Astropy Collaboration} {et~al.}(2013){Astropy Collaboration}, {Robitaille}, {Tollerud}, {Greenfield}, {Droettboom}, {Bray}, {Aldcroft}, {Davis}, {Ginsburg}, {Price-Whelan}, {Kerzendorf}, {Conley}, {Crighton}, {Barbary}, {Muna}, {Ferguson}, {Grollier}, {Parikh}, {Nair}, {Unther}, {Deil}, {Woillez}, {Conseil}, {Kramer}, {Turner}, {Singer}, {Fox}, {Weaver}, {Zabalza}, {Edwards}, {Azalee Bostroem}, {Burke}, {Casey}, {Crawford}, {Dencheva}, {Ely}, {Jenness}, {Labrie}, {Lim}, {Pierfederici}, {Pontzen}, {Ptak}, {Refsdal}, {Servillat}, \& {Streicher}}]{astropy}
{Astropy Collaboration}, {Robitaille}, T.~P., {Tollerud}, E.~J., {et~al.} 2013, \aap, 558, A33, \dodoi{10.1051/0004-6361/201322068}

\bibitem[{{Avenhaus} {et~al.}(2018){Avenhaus}, {Quanz}, {Garufi}, {Perez}, {Casassus}, {Pinte}, {Bertrang}, {Caceres}, {Benisty}, \& {Dominik}}]{aven18}
{Avenhaus}, H., {Quanz}, S.~P., {Garufi}, A., {et~al.} 2018, \apj, 863, 44, \dodoi{10.3847/1538-4357/aab846}

\bibitem[{{Barrado Y Navascu{\'e}s}(2006)}]{barr06}
{Barrado Y Navascu{\'e}s}, D. 2006, \aap, 459, 511, \dodoi{10.1051/0004-6361:20065717}

\bibitem[{{Bergin} {et~al.}(2013){Bergin}, {Cleeves}, {Gorti}, {Zhang}, {Blake}, {Green}, {Andrews}, {Evans}, {Henning}, {{\"O}berg}, {Pontoppidan}, {Qi}, {Salyk}, \& {van Dishoeck}}]{berg13}
{Bergin}, E.~A., {Cleeves}, L.~I., {Gorti}, U., {et~al.} 2013, \nat, 493, 644, \dodoi{10.1038/nature11805}

\bibitem[{{Birnstiel} {et~al.}(2018){Birnstiel}, {Dullemond}, {Zhu}, {Andrews}, {Bai}, {Wilner}, {Carpenter}, {Huang}, {Isella}, {Benisty}, {P{\'e}rez}, \& {Zhang}}]{birn18}
{Birnstiel}, T., {Dullemond}, C.~P., {Zhu}, Z., {et~al.} 2018, \apjl, 869, L45, \dodoi{10.3847/2041-8213/aaf743}

\bibitem[{{Bohren} \& {Huffman}(1998)}]{bohr98}
{Bohren}, C.~F., \& {Huffman}, D.~R. 1998, {Absorption and Scattering of Light by Small Particles}

\bibitem[{{Bosman} \& {Banzatti}(2019)}]{bosm19}
{Bosman}, A.~D., \& {Banzatti}, A. 2019, \aap, 632, L10, \dodoi{10.1051/0004-6361/201936638}

\bibitem[{{Bosman} {et~al.}(2021){Bosman}, {Bergin}, {Loomis}, {Andrews}, {van't Hoff}, {Teague}, {{\"O}berg}, {Guzm{\'a}n}, {Walsh}, {Aikawa}, {Alarc{\'o}n}, {Bae}, {Bergner}, {Booth}, {Cataldi}, {Cleeves}, {Czekala}, {Huang}, {Ilee}, {Law}, {Le Gal}, {Liu}, {Long}, {M{\'e}nard}, {Nomura}, {P{\'e}rez}, {Qi}, {Schwarz}, {Sierra}, {Tsukagoshi}, {Yamato}, {Wilner}, \& {Zhang}}]{bosm21}
{Bosman}, A.~D., {Bergin}, E.~A., {Loomis}, R.~A., {et~al.} 2021, \apjs, 257, 15, \dodoi{10.3847/1538-4365/ac1433}

\bibitem[{{Brogan} {et~al.}(2018){Brogan}, {Hunter}, \& {Fomalont}}]{brog18}
{Brogan}, C.~L., {Hunter}, T.~R., \& {Fomalont}, E.~B. 2018, arXiv e-prints, arXiv:1805.05266, \dodoi{10.48550/arXiv.1805.05266}

\bibitem[{{Canta} {et~al.}(2021){Canta}, {Teague}, {Le Gal}, \& {{\"O}berg}}]{cant21}
{Canta}, A., {Teague}, R., {Le Gal}, R., \& {{\"O}berg}, K.~I. 2021, \apj, 922, 62, \dodoi{10.3847/1538-4357/ac23da}

\bibitem[{{Carrasco-Gonz{\'a}lez} {et~al.}(2019){Carrasco-Gonz{\'a}lez}, {Sierra}, {Flock}, {Zhu}, {Henning}, {Chandler}, {Galv{\'a}n-Madrid}, {Mac{\'\i}as}, {Anglada}, {Linz}, {Osorio}, {Rodr{\'\i}guez}, {Testi}, {Torrelles}, {P{\'e}rez}, \& {Liu}}]{carr19}
{Carrasco-Gonz{\'a}lez}, C., {Sierra}, A., {Flock}, M., {et~al.} 2019, \apj, 883, 71, \dodoi{10.3847/1538-4357/ab3d33}

\bibitem[{{Casassus} \& {C{\'a}rcamo}(2022)}]{casa22}
{Casassus}, S., \& {C{\'a}rcamo}, M. 2022, \mnras, 513, 5790, \dodoi{10.1093/mnras/stac1285}

\bibitem[{{Czekala} {et~al.}(2021){Czekala}, {Loomis}, {Teague}, {Booth}, {Huang}, {Cataldi}, {Ilee}, {Law}, {Walsh}, {Bosman}, {Guzm{\'a}n}, {Gal}, {{\"O}berg}, {Yamato}, {Aikawa}, {Andrews}, {Bae}, {Bergin}, {Bergner}, {Cleeves}, {Kurtovic}, {M{\'e}nard}, {Nomura}, {P{\'e}rez}, {Qi}, {Schwarz}, {Tsukagoshi}, {Waggoner}, {Wilner}, \& {Zhang}}]{czek21}
{Czekala}, I., {Loomis}, R.~A., {Teague}, R., {et~al.} 2021, \apjs, 257, 2, \dodoi{10.3847/1538-4365/ac1430}

\bibitem[{{Delussu} {et~al.}(2024){Delussu}, {Birnstiel}, {Miotello}, {Pinilla}, {Rosotti}, \& {Andrews}}]{delu24}
{Delussu}, L., {Birnstiel}, T., {Miotello}, A., {et~al.} 2024, \aap, 688, A81, \dodoi{10.1051/0004-6361/202450328}

\bibitem[{{Doi} \& {Kataoka}(2023)}]{doi23}
{Doi}, K., \& {Kataoka}, A. 2023, \apj, 957, 11, \dodoi{10.3847/1538-4357/acf5df}

\bibitem[{{Dominik} \& {Dullemond}(2024)}]{domi24}
{Dominik}, C., \& {Dullemond}, C.~P. 2024, \aap, 682, A144, \dodoi{10.1051/0004-6361/202347716}

\bibitem[{{Dominik} {et~al.}(2021){Dominik}, {Min}, \& {Tazaki}}]{optool}
{Dominik}, C., {Min}, M., \& {Tazaki}, R. 2021, {OpTool: Command-line driven tool for creating complex dust opacities}, Astrophysics Source Code Library, record ascl:2104.010.
\newblock \doeprint{2104.010}

\bibitem[{{Dorschner} {et~al.}(1995){Dorschner}, {Begemann}, {Henning}, {Jaeger}, \& {Mutschke}}]{dors95}
{Dorschner}, J., {Begemann}, B., {Henning}, T., {Jaeger}, C., \& {Mutschke}, H. 1995, \aap, 300, 503

\bibitem[{{Draine}(2003)}]{drai03}
{Draine}, B.~T. 2003, \araa, 41, 241, \dodoi{10.1146/annurev.astro.41.011802.094840}

\bibitem[{{Dubrulle} {et~al.}(1995){Dubrulle}, {Morfill}, \& {Sterzik}}]{dubr95}
{Dubrulle}, B., {Morfill}, G., \& {Sterzik}, M. 1995, \icarus, 114, 237, \dodoi{10.1006/icar.1995.1058}

\bibitem[{{Dullemond} {et~al.}(2012){Dullemond}, {Juhasz}, {Pohl}, {Sereshti}, {Shetty}, {Peters}, {Commercon}, \& {Flock}}]{dull12}
{Dullemond}, C.~P., {Juhasz}, A., {Pohl}, A., {et~al.} 2012, {RADMC-3D: A multi-purpose radiative transfer tool}, Astrophysics Source Code Library, record ascl:1202.015.
\newblock \doeprint{1202.015}

\bibitem[{{Dullemond} {et~al.}(2002){Dullemond}, {van Zadelhoff}, \& {Natta}}]{dull02}
{Dullemond}, C.~P., {van Zadelhoff}, G.~J., \& {Natta}, A. 2002, \aap, 389, 464, \dodoi{10.1051/0004-6361:20020608}

\bibitem[{{Foreman-Mackey} {et~al.}(2013){Foreman-Mackey}, {Hogg}, {Lang}, \& {Goodman}}]{emcee}
{Foreman-Mackey}, D., {Hogg}, D.~W., {Lang}, D., \& {Goodman}, J. 2013, \pasp, 125, 306, \dodoi{10.1086/670067}

\bibitem[{{Francis} {et~al.}(2020){Francis}, {Johnstone}, {Herczeg}, {Hunter}, \& {Harsono}}]{fran20}
{Francis}, L., {Johnstone}, D., {Herczeg}, G., {Hunter}, T.~R., \& {Harsono}, D. 2020, \aj, 160, 270, \dodoi{10.3847/1538-3881/abbe1a}

\bibitem[{{Gaia Collaboration} {et~al.}(2016){Gaia Collaboration}, {Prusti}, {de Bruijne}, {Brown}, {Vallenari}, {Babusiaux}, {Bailer-Jones}, {Bastian}, {Biermann}, {Evans}, {Eyer}, {Jansen}, {Jordi}, {Klioner}, {Lammers}, {Lindegren}, {Luri}, {Mignard}, {Milligan}, {Panem}, {Poinsignon}, {Pourbaix}, {Randich}, {Sarri}, {Sartoretti}, {Siddiqui}, {Soubiran}, {Valette}, {van Leeuwen}, {Walton}, {Aerts}, {Arenou}, {Cropper}, {Drimmel}, {H{\o}g}, {Katz}, {Lattanzi}, {O'Mullane}, {Grebel}, {Holland}, {Huc}, {Passot}, {Bramante}, {Cacciari}, {Casta{\~n}eda}, {Chaoul}, {Cheek}, {De Angeli}, {Fabricius}, {Guerra}, {Hern{\'a}ndez}, {Jean-Antoine-Piccolo}, {Masana}, {Messineo}, {Mowlavi}, {Nienartowicz}, {Ord{\'o}{\~n}ez-Blanco}, {Panuzzo}, {Portell}, {Richards}, {Riello}, {Seabroke}, {Tanga}, {Th{\'e}venin}, {Torra}, {Els}, {Gracia-Abril}, {Comoretto}, {Garcia-Reinaldos}, {Lock}, {Mercier}, {Altmann}, {Andrae}, {Astraatmadja}, {Bellas-Velidis}, {Benson}, {Berthier}, {Blomme}, {Busso}, {Carry}, {Cellino}, {Clementini},
  {Cowell}, {Creevey}, {Cuypers}, {Davidson}, {De Ridder}, {de Torres}, {Delchambre}, {Dell'Oro}, {Ducourant}, {Fr{\'e}mat}, {Garc{\'\i}a-Torres}, {Gosset}, {Halbwachs}, {Hambly}, {Harrison}, {Hauser}, {Hestroffer}, {Hodgkin}, {Huckle}, {Hutton}, {Jasniewicz}, {Jordan}, {Kontizas}, {Korn}, {Lanzafame}, {Manteiga}, {Moitinho}, {Muinonen}, {Osinde}, {Pancino}, {Pauwels}, {Petit}, {Recio-Blanco}, {Robin}, {Sarro}, {Siopis}, {Smith}, {Smith}, {Sozzetti}, {Thuillot}, {van Reeven}, {Viala}, {Abbas}, {Abreu Aramburu}, {Accart}, {Aguado}, {Allan}, {Allasia}, {Altavilla}, {{\'A}lvarez}, {Alves}, {Anderson}, {Andrei}, {Anglada Varela}, {Antiche}, {Antoja}, {Ant{\'o}n}, {Arcay}, {Atzei}, {Ayache}, {Bach}, {Baker}, {Balaguer-N{\'u}{\~n}ez}, {Barache}, {Barata}, {Barbier}, {Barblan}, {Baroni}, {Barrado y Navascu{\'e}s}, {Barros}, {Barstow}, {Becciani}, {Bellazzini}, {Bellei}, {Bello Garc{\'\i}a}, {Belokurov}, {Bendjoya}, {Berihuete}, {Bianchi}, {Bienaym{\'e}}, {Billebaud}, {Blagorodnova}, {Blanco-Cuaresma}, {Boch},
  {Bombrun}, {Borrachero}, {Bouquillon}, {Bourda}, {Bouy}, {Bragaglia}, {Breddels}, {Brouillet}, {Br{\"u}semeister}, {Bucciarelli}, {Budnik}, {Burgess}, {Burgon}, {Burlacu}, {Busonero}, {Buzzi}, {Caffau}, {Cambras}, {Campbell}, {Cancelliere}, {Cantat-Gaudin}, {Carlucci}, {Carrasco}, {Castellani}, {Charlot}, {Charnas}, {Charvet}, {Chassat}, {Chiavassa}, {Clotet}, {Cocozza}, {Collins}, {Collins}, {Costigan}, {Crifo}, {Cross}, {Crosta}, {Crowley}, {Dafonte}, {Damerdji}, {Dapergolas}, {David}, {David}, {De Cat}, {de Felice}, {de Laverny}, {De Luise}, {De March}, {de Martino}, {de Souza}, {Debosscher}, {del Pozo}, {Delbo}, {Delgado}, {Delgado}, {di Marco}, {Di Matteo}, {Diakite}, {Distefano}, {Dolding}, {Dos Anjos}, {Drazinos}, {Dur{\'a}n}, {Dzigan}, {Ecale}, {Edvardsson}, {Enke}, {Erdmann}, {Escolar}, {Espina}, {Evans}, {Eynard Bontemps}, {Fabre}, {Fabrizio}, {Faigler}, {Falc{\~a}o}, {Farr{\`a}s Casas}, {Faye}, {Federici}, {Fedorets}, {Fern{\'a}ndez-Hern{\'a}ndez}, {Fernique}, {Fienga}, {Figueras}, {Filippi},
  {Findeisen}, {Fonti}, {Fouesneau}, {Fraile}, {Fraser}, {Fuchs}, {Furnell}, {Gai}, {Galleti}, {Galluccio}, {Garabato}, {Garc{\'\i}a-Sedano}, {Gar{\'e}}, {Garofalo}, {Garralda}, {Gavras}, {Gerssen}, {Geyer}, {Gilmore}, {Girona}, {Giuffrida}, {Gomes}, {Gonz{\'a}lez-Marcos}, {Gonz{\'a}lez-N{\'u}{\~n}ez}, {Gonz{\'a}lez-Vidal}, {Granvik}, {Guerrier}, {Guillout}, {Guiraud}, {G{\'u}rpide}, {Guti{\'e}rrez-S{\'a}nchez}, {Guy}, {Haigron}, {Hatzidimitriou}, {Haywood}, {Heiter}, {Helmi}, {Hobbs}, {Hofmann}, {Holl}, {Holland}, {Hunt}, {Hypki}, {Icardi}, {Irwin}, {Jevardat de Fombelle}, {Jofr{\'e}}, {Jonker}, {Jorissen}, {Julbe}, {Karampelas}, {Kochoska}, {Kohley}, {Kolenberg}, {Kontizas}, {Koposov}, {Kordopatis}, {Koubsky}, {Kowalczyk}, {Krone-Martins}, {Kudryashova}, {Kull}, {Bachchan}, {Lacoste-Seris}, {Lanza}, {Lavigne}, {Le Poncin-Lafitte}, {Lebreton}, {Lebzelter}, {Leccia}, {Leclerc}, {Lecoeur-Taibi}, {Lemaitre}, {Lenhardt}, {Leroux}, {Liao}, {Licata}, {Lindstr{\o}m}, {Lister}, {Livanou}, {Lobel}, {L{\"o}ffler},
  {L{\'o}pez}, {Lopez-Lozano}, {Lorenz}, {Loureiro}, {MacDonald}, {Magalh{\~a}es Fernandes}, {Managau}, {Mann}, {Mantelet}, {Marchal}, {Marchant}, {Marconi}, {Marie}, {Marinoni}, {Marrese}, {Marschalk{\'o}}, {Marshall}, {Mart{\'\i}n-Fleitas}, {Martino}, {Mary}, {Matijevi{\v{c}}}, {Mazeh}, {McMillan}, {Messina}, {Mestre}, {Michalik}, {Millar}, {Miranda}, {Molina}, {Molinaro}, {Molinaro}, {Moln{\'a}r}, {Moniez}, {Montegriffo}, {Monteiro}, {Mor}, {Mora}, {Morbidelli}, {Morel}, {Morgenthaler}, {Morley}, {Morris}, {Mulone}, {Muraveva}, {Musella}, {Narbonne}, {Nelemans}, {Nicastro}, {Noval}, {Ord{\'e}novic}, {Ordieres-Mer{\'e}}, {Osborne}, {Pagani}, {Pagano}, {Pailler}, {Palacin}, {Palaversa}, {Parsons}, {Paulsen}, {Pecoraro}, {Pedrosa}, {Pentik{\"a}inen}, {Pereira}, {Pichon}, {Piersimoni}, {Pineau}, {Plachy}, {Plum}, {Poujoulet}, {Pr{\v{s}}a}, {Pulone}, {Ragaini}, {Rago}, {Rambaux}, {Ramos-Lerate}, {Ranalli}, {Rauw}, {Read}, {Regibo}, {Renk}, {Reyl{\'e}}, {Ribeiro}, {Rimoldini}, {Ripepi}, {Riva}, {Rixon},
  {Roelens}, {Romero-G{\'o}mez}, {Rowell}, {Royer}, {Rudolph}, {Ruiz-Dern}, {Sadowski}, {Sagrist{\`a} Sell{\'e}s}, {Sahlmann}, {Salgado}, {Salguero}, {Sarasso}, {Savietto}, {Schnorhk}, {Schultheis}, {Sciacca}, {Segol}, {Segovia}, {Segransan}, {Serpell}, {Shih}, {Smareglia}, {Smart}, {Smith}, {Solano}, {Solitro}, {Sordo}, {Soria Nieto}, {Souchay}, {Spagna}, {Spoto}, {Stampa}, {Steele}, {Steidelm{\"u}ller}, {Stephenson}, {Stoev}, {Suess}, {S{\"u}veges}, {Surdej}, {Szabados}, {Szegedi-Elek}, {Tapiador}, {Taris}, {Tauran}, {Taylor}, {Teixeira}, {Terrett}, {Tingley}, {Trager}, {Turon}, {Ulla}, {Utrilla}, {Valentini}, {van Elteren}, {Van Hemelryck}, {van Leeuwen}, {Varadi}, {Vecchiato}, {Veljanoski}, {Via}, {Vicente}, {Vogt}, {Voss}, {Votruba}, {Voutsinas}, {Walmsley}, {Weiler}, {Weingrill}, {Werner}, {Wevers}, {Whitehead}, {Wyrzykowski}, {Yoldas}, {{\v{Z}}erjal}, {Zucker}, {Zurbach}, {Zwitter}, {Alecu}, {Allen}, {Allende Prieto}, {Amorim}, {Anglada-Escud{\'e}}, {Arsenijevic}, {Azaz}, {Balm}, {Beck}, {Bernstein},
  {Bigot}, {Bijaoui}, {Blasco}, {Bonfigli}, {Bono}, {Boudreault}, {Bressan}, {Brown}, {Brunet}, {Bunclark}, {Buonanno}, {Butkevich}, {Carret}, {Carrion}, {Chemin}, {Ch{\'e}reau}, {Corcione}, {Darmigny}, {de Boer}, {de Teodoro}, {de Zeeuw}, {Delle Luche}, {Domingues}, {Dubath}, {Fodor}, {Fr{\'e}zouls}, {Fries}, {Fustes}, {Fyfe}, {Gallardo}, {Gallegos}, {Gardiol}, {Gebran}, {Gomboc}, {G{\'o}mez}, {Grux}, {Gueguen}, {Heyrovsky}, {Hoar}, {Iannicola}, {Isasi Parache}, {Janotto}, {Joliet}, {Jonckheere}, {Keil}, {Kim}, {Klagyivik}, {Klar}, {Knude}, {Kochukhov}, {Kolka}, {Kos}, {Kutka}, {Lainey}, {LeBouquin}, {Liu}, {Loreggia}, {Makarov}, {Marseille}, {Martayan}, {Martinez-Rubi}, {Massart}, {Meynadier}, {Mignot}, {Munari}, {Nguyen}, {Nordlander}, {Ocvirk}, {O'Flaherty}, {Olias Sanz}, {Ortiz}, {Osorio}, {Oszkiewicz}, {Ouzounis}, {Palmer}, {Park}, {Pasquato}, {Peltzer}, {Peralta}, {P{\'e}turaud}, {Pieniluoma}, {Pigozzi}, {Poels}, {Prat}, {Prod'homme}, {Raison}, {Rebordao}, {Risquez}, {Rocca-Volmerange}, {Rosen},
  {Ruiz-Fuertes}, {Russo}, {Sembay}, {Serraller Vizcaino}, {Short}, {Siebert}, {Silva}, {Sinachopoulos}, {Slezak}, {Soffel}, {Sosnowska}, {Strai{\v{z}}ys}, {ter Linden}, {Terrell}, {Theil}, {Tiede}, {Troisi}, {Tsalmantza}, {Tur}, {Vaccari}, {Vachier}, {Valles}, {Van Hamme}, {Veltz}, {Virtanen}, {Wallut}, {Wichmann}, {Wilkinson}, {Ziaeepour}, \& {Zschocke}}]{gaia16}
{Gaia Collaboration}, {Prusti}, T., {de Bruijne}, J.~H.~J., {et~al.} 2016, \aap, 595, A1, \dodoi{10.1051/0004-6361/201629272}

\bibitem[{{Gaia Collaboration} {et~al.}(2021){Gaia Collaboration}, {Brown}, {Vallenari}, {Prusti}, {de Bruijne}, {Babusiaux}, {Biermann}, {Creevey}, {Evans}, {Eyer}, {Hutton}, {Jansen}, {Jordi}, {Klioner}, {Lammers}, {Lindegren}, {Luri}, {Mignard}, {Panem}, {Pourbaix}, {Randich}, {Sartoretti}, {Soubiran}, {Walton}, {Arenou}, {Bailer-Jones}, {Bastian}, {Cropper}, {Drimmel}, {Katz}, {Lattanzi}, {van Leeuwen}, {Bakker}, {Cacciari}, {Casta{\~n}eda}, {De Angeli}, {Ducourant}, {Fabricius}, {Fouesneau}, {Fr{\'e}mat}, {Guerra}, {Guerrier}, {Guiraud}, {Jean-Antoine Piccolo}, {Masana}, {Messineo}, {Mowlavi}, {Nicolas}, {Nienartowicz}, {Pailler}, {Panuzzo}, {Riclet}, {Roux}, {Seabroke}, {Sordo}, {Tanga}, {Th{\'e}venin}, {Gracia-Abril}, {Portell}, {Teyssier}, {Altmann}, {Andrae}, {Bellas-Velidis}, {Benson}, {Berthier}, {Blomme}, {Brugaletta}, {Burgess}, {Busso}, {Carry}, {Cellino}, {Cheek}, {Clementini}, {Damerdji}, {Davidson}, {Delchambre}, {Dell'Oro}, {Fern{\'a}ndez-Hern{\'a}ndez}, {Galluccio}, {Garc{\'\i}a-Lario},
  {Garcia-Reinaldos}, {Gonz{\'a}lez-N{\'u}{\~n}ez}, {Gosset}, {Haigron}, {Halbwachs}, {Hambly}, {Harrison}, {Hatzidimitriou}, {Heiter}, {Hern{\'a}ndez}, {Hestroffer}, {Hodgkin}, {Holl}, {Jan{\ss}en}, {Jevardat de Fombelle}, {Jordan}, {Krone-Martins}, {Lanzafame}, {L{\"o}ffler}, {Lorca}, {Manteiga}, {Marchal}, {Marrese}, {Moitinho}, {Mora}, {Muinonen}, {Osborne}, {Pancino}, {Pauwels}, {Petit}, {Recio-Blanco}, {Richards}, {Riello}, {Rimoldini}, {Robin}, {Roegiers}, {Rybizki}, {Sarro}, {Siopis}, {Smith}, {Sozzetti}, {Ulla}, {Utrilla}, {van Leeuwen}, {van Reeven}, {Abbas}, {Abreu Aramburu}, {Accart}, {Aerts}, {Aguado}, {Ajaj}, {Altavilla}, {{\'A}lvarez}, {{\'A}lvarez Cid-Fuentes}, {Alves}, {Anderson}, {Anglada Varela}, {Antoja}, {Audard}, {Baines}, {Baker}, {Balaguer-N{\'u}{\~n}ez}, {Balbinot}, {Balog}, {Barache}, {Barbato}, {Barros}, {Barstow}, {Bartolom{\'e}}, {Bassilana}, {Bauchet}, {Baudesson-Stella}, {Becciani}, {Bellazzini}, {Bernet}, {Bertone}, {Bianchi}, {Blanco-Cuaresma}, {Boch}, {Bombrun}, {Bossini},
  {Bouquillon}, {Bragaglia}, {Bramante}, {Breedt}, {Bressan}, {Brouillet}, {Bucciarelli}, {Burlacu}, {Busonero}, {Butkevich}, {Buzzi}, {Caffau}, {Cancelliere}, {C{\'a}novas}, {Cantat-Gaudin}, {Carballo}, {Carlucci}, {Carnerero}, {Carrasco}, {Casamiquela}, {Castellani}, {Castro-Ginard}, {Castro Sampol}, {Chaoul}, {Charlot}, {Chemin}, {Chiavassa}, {Cioni}, {Comoretto}, {Cooper}, {Cornez}, {Cowell}, {Crifo}, {Crosta}, {Crowley}, {Dafonte}, {Dapergolas}, {David}, {David}, {de Laverny}, {De Luise}, {De March}, {De Ridder}, {de Souza}, {de Teodoro}, {de Torres}, {del Peloso}, {del Pozo}, {Delbo}, {Delgado}, {Delgado}, {Delisle}, {Di Matteo}, {Diakite}, {Diener}, {Distefano}, {Dolding}, {Eappachen}, {Edvardsson}, {Enke}, {Esquej}, {Fabre}, {Fabrizio}, {Faigler}, {Fedorets}, {Fernique}, {Fienga}, {Figueras}, {Fouron}, {Fragkoudi}, {Fraile}, {Franke}, {Gai}, {Garabato}, {Garcia-Gutierrez}, {Garc{\'\i}a-Torres}, {Garofalo}, {Gavras}, {Gerlach}, {Geyer}, {Giacobbe}, {Gilmore}, {Girona}, {Giuffrida}, {Gomel}, {Gomez},
  {Gonzalez-Santamaria}, {Gonz{\'a}lez-Vidal}, {Granvik}, {Guti{\'e}rrez-S{\'a}nchez}, {Guy}, {Hauser}, {Haywood}, {Helmi}, {Hidalgo}, {Hilger}, {H{\l}adczuk}, {Hobbs}, {Holland}, {Huckle}, {Jasniewicz}, {Jonker}, {Juaristi Campillo}, {Julbe}, {Karbevska}, {Kervella}, {Khanna}, {Kochoska}, {Kontizas}, {Kordopatis}, {Korn}, {Kostrzewa-Rutkowska}, {Kruszy{\'n}ska}, {Lambert}, {Lanza}, {Lasne}, {Le Campion}, {Le Fustec}, {Lebreton}, {Lebzelter}, {Leccia}, {Leclerc}, {Lecoeur-Taibi}, {Liao}, {Licata}, {Lindstr{\o}m}, {Lister}, {Livanou}, {Lobel}, {Madrero Pardo}, {Managau}, {Mann}, {Marchant}, {Marconi}, {Marcos Santos}, {Marinoni}, {Marocco}, {Marshall}, {Martin Polo}, {Mart{\'\i}n-Fleitas}, {Masip}, {Massari}, {Mastrobuono-Battisti}, {Mazeh}, {McMillan}, {Messina}, {Michalik}, {Millar}, {Mints}, {Molina}, {Molinaro}, {Moln{\'a}r}, {Montegriffo}, {Mor}, {Morbidelli}, {Morel}, {Morris}, {Mulone}, {Munoz}, {Muraveva}, {Murphy}, {Musella}, {Noval}, {Ord{\'e}novic}, {Orr{\`u}}, {Osinde}, {Pagani}, {Pagano},
  {Palaversa}, {Palicio}, {Panahi}, {Pawlak}, {Pe{\~n}alosa Esteller}, {Penttil{\"a}}, {Piersimoni}, {Pineau}, {Plachy}, {Plum}, {Poggio}, {Poretti}, {Poujoulet}, {Pr{\v{s}}a}, {Pulone}, {Racero}, {Ragaini}, {Rainer}, {Raiteri}, {Rambaux}, {Ramos}, {Ramos-Lerate}, {Re Fiorentin}, {Regibo}, {Reyl{\'e}}, {Ripepi}, {Riva}, {Rixon}, {Robichon}, {Robin}, {Roelens}, {Rohrbasser}, {Romero-G{\'o}mez}, {Rowell}, {Royer}, {Rybicki}, {Sadowski}, {Sagrist{\`a} Sell{\'e}s}, {Sahlmann}, {Salgado}, {Salguero}, {Samaras}, {Sanchez Gimenez}, {Sanna}, {Santove{\~n}a}, {Sarasso}, {Schultheis}, {Sciacca}, {Segol}, {Segovia}, {S{\'e}gransan}, {Semeux}, {Shahaf}, {Siddiqui}, {Siebert}, {Siltala}, {Slezak}, {Smart}, {Solano}, {Solitro}, {Souami}, {Souchay}, {Spagna}, {Spoto}, {Steele}, {Steidelm{\"u}ller}, {Stephenson}, {S{\"u}veges}, {Szabados}, {Szegedi-Elek}, {Taris}, {Tauran}, {Taylor}, {Teixeira}, {Thuillot}, {Tonello}, {Torra}, {Torra}, {Turon}, {Unger}, {Vaillant}, {van Dillen}, {Vanel}, {Vecchiato}, {Viala}, {Vicente},
  {Voutsinas}, {Weiler}, {Wevers}, {Wyrzykowski}, {Yoldas}, {Yvard}, {Zhao}, {Zorec}, {Zucker}, {Zurbach}, \& {Zwitter}}]{gaia21}
{Gaia Collaboration}, {Brown}, A.~G.~A., {Vallenari}, A., {et~al.} 2021, \aap, 649, A1, \dodoi{10.1051/0004-6361/202039657}

\bibitem[{Gordon {et~al.}(2022)Gordon, Rothman, Hargreaves, Hashemi, Karlovets, Skinner, Conway, Hill, Kochanov, Tan, Wcisło, Finenko, Nelson, Bernath, Birk, Boudon, Campargue, Chance, Coustenis, Drouin, Flaud, Gamache, Hodges, Jacquemart, Mlawer, Nikitin, Perevalov, Rotger, Tennyson, Toon, Tran, Tyuterev, Adkins, Baker, Barbe, Canè, Császár, Dudaryonok, Egorov, Fleisher, Fleurbaey, Foltynowicz, Furtenbacher, Harrison, Hartmann, Horneman, Huang, Karman, Karns, Kassi, Kleiner, Kofman, Kwabia–Tchana, Lavrentieva, Lee, Long, Lukashevskaya, Lyulin, Makhnev, Matt, Massie, Melosso, Mikhailenko, Mondelain, Müller, Naumenko, Perrin, Polyansky, Raddaoui, Raston, Reed, Rey, Richard, Tóbiás, Sadiek, Schwenke, Starikova, Sung, Tamassia, Tashkun, {Vander Auwera}, Vasilenko, Vigasin, Villanueva, Vispoel, Wagner, Yachmenev, \& Yurchenko}]{hitran}
Gordon, I., Rothman, L., Hargreaves, R., {et~al.} 2022, Journal of Quantitative Spectroscopy and Radiative Transfer, 277, 107949, \dodoi{https://doi.org/10.1016/j.jqsrt.2021.107949}

\bibitem[{{Henning} \& {Stognienko}(1996)}]{henn96}
{Henning}, T., \& {Stognienko}, R. 1996, \aap, 311, 291

\bibitem[{{Inoue} {et~al.}(2009){Inoue}, {Oka}, \& {Nakamoto}}]{inou09}
{Inoue}, A.~K., {Oka}, A., \& {Nakamoto}, T. 2009, \mnras, 393, 1377, \dodoi{10.1111/j.1365-2966.2008.14316.x}

\bibitem[{{Ishimaru}(1978)}]{ishi78}
{Ishimaru}, A. 1978, {Wave propagation and scattering in random media. Volume 1 - Single scattering and transport theory}, Vol.~1, \dodoi{10.1016/B978-0-12-374701-3.X5001-7}

\bibitem[{{Jorsater} \& {van Moorsel}(1995)}]{jors95}
{Jorsater}, S., \& {van Moorsel}, G.~A. 1995, \aj, 110, 2037, \dodoi{10.1086/117668}

\bibitem[{{Kozasa} {et~al.}(1992){Kozasa}, {Blum}, \& {Mukai}}]{koza92}
{Kozasa}, T., {Blum}, J., \& {Mukai}, T. 1992, \aap, 263, 423

\bibitem[{{Krijt} {et~al.}(2020){Krijt}, {Bosman}, {Zhang}, {Schwarz}, {Ciesla}, \& {Bergin}}]{krij20}
{Krijt}, S., {Bosman}, A.~D., {Zhang}, K., {et~al.} 2020, \apj, 899, 134, \dodoi{10.3847/1538-4357/aba75d}

\bibitem[{{Mac{\'\i}as} {et~al.}(2021){Mac{\'\i}as}, {Guerra-Alvarado}, {Carrasco-Gonz{\'a}lez}, {Ribas}, {Espaillat}, {Huang}, \& {Andrews}}]{maci21}
{Mac{\'\i}as}, E., {Guerra-Alvarado}, O., {Carrasco-Gonz{\'a}lez}, C., {et~al.} 2021, \aap, 648, A33, \dodoi{10.1051/0004-6361/202039812}

\bibitem[{{Mathis} {et~al.}(1977){Mathis}, {Rumpl}, \& {Nordsieck}}]{math77}
{Mathis}, J.~S., {Rumpl}, W., \& {Nordsieck}, K.~H. 1977, \apj, 217, 425, \dodoi{10.1086/155591}

\bibitem[{{McMullin} {et~al.}(2007){McMullin}, {Waters}, {Schiebel}, {Young}, \& {Golap}}]{mcmu07}
{McMullin}, J.~P., {Waters}, B., {Schiebel}, D., {Young}, W., \& {Golap}, K. 2007, in Astronomical Society of the Pacific Conference Series, Vol. 376, Astronomical Data Analysis Software and Systems XVI, ed. R.~A. {Shaw}, F.~{Hill}, \& D.~J. {Bell}, 127

\bibitem[{{Miyake} \& {Nakagawa}(1993)}]{miya93}
{Miyake}, K., \& {Nakagawa}, Y. 1993, \icarus, 106, 20, \dodoi{10.1006/icar.1993.1156}

\bibitem[{{Mukai} {et~al.}(1992){Mukai}, {Ishimoto}, {Kozasa}, {Blum}, \& {Greenberg}}]{muka92}
{Mukai}, T., {Ishimoto}, H., {Kozasa}, T., {Blum}, J., \& {Greenberg}, J.~M. 1992, \aap, 262, 315

\bibitem[{{Musiolik} {et~al.}(2016){Musiolik}, {Teiser}, {Jankowski}, \& {Wurm}}]{musi16}
{Musiolik}, G., {Teiser}, J., {Jankowski}, T., \& {Wurm}, G. 2016, \apj, 827, 63, \dodoi{10.3847/0004-637X/827/1/63}

\bibitem[{{{\"O}berg} \& {Bergin}(2021)}]{ober21}
{{\"O}berg}, K.~I., \& {Bergin}, E.~A. 2021, \physrep, 893, 1, \dodoi{10.1016/j.physrep.2020.09.004}

\bibitem[{{Okuzumi} {et~al.}(2012){Okuzumi}, {Tanaka}, {Kobayashi}, \& {Wada}}]{okuz12}
{Okuzumi}, S., {Tanaka}, H., {Kobayashi}, H., \& {Wada}, K. 2012, \apj, 752, 106, \dodoi{10.1088/0004-637X/752/2/106}

\bibitem[{{Okuzumi} \& {Tazaki}(2019)}]{okuz19}
{Okuzumi}, S., \& {Tazaki}, R. 2019, \apj, 878, 132, \dodoi{10.3847/1538-4357/ab204d}

\bibitem[{{Pinte} {et~al.}(2009){Pinte}, {Harries}, {Min}, {Watson}, {Dullemond}, {Woitke}, {M{\'e}nard}, \& {Dur{\'a}n-Rojas}}]{pint09}
{Pinte}, C., {Harries}, T.~J., {Min}, M., {et~al.} 2009, \aap, 498, 967, \dodoi{10.1051/0004-6361/200811555}

\bibitem[{{Ricci} {et~al.}(2010){Ricci}, {Testi}, {Natta}, {Neri}, {Cabrit}, \& {Herczeg}}]{ricc10}
{Ricci}, L., {Testi}, L., {Natta}, A., {et~al.} 2010, \aap, 512, A15, \dodoi{10.1051/0004-6361/200913403}

\bibitem[{{Rybicki} \& {Lightman}(1979)}]{rybi79}
{Rybicki}, G.~B., \& {Lightman}, A.~P. 1979, {Radiative processes in astrophysics}

\bibitem[{{Sch{\"o}ier} {et~al.}(2005){Sch{\"o}ier}, {van der Tak}, {van Dishoeck}, \& {Black}}]{lamda}
{Sch{\"o}ier}, F.~L., {van der Tak}, F.~F.~S., {van Dishoeck}, E.~F., \& {Black}, J.~H. 2005, \aap, 432, 369, \dodoi{10.1051/0004-6361:20041729}

\bibitem[{{Shakura} \& {Sunyaev}(1973)}]{shak73}
{Shakura}, N.~I., \& {Sunyaev}, R.~A. 1973, \aap, 24, 337

\bibitem[{{Shen} {et~al.}(2008){Shen}, {Draine}, \& {Johnson}}]{shen08}
{Shen}, Y., {Draine}, B.~T., \& {Johnson}, E.~T. 2008, \apj, 689, 260, \dodoi{10.1086/592765}

\bibitem[{{Tazaki} {et~al.}(2019){Tazaki}, {Tanaka}, {Kataoka}, {Okuzumi}, \& {Muto}}]{taza19}
{Tazaki}, R., {Tanaka}, H., {Kataoka}, A., {Okuzumi}, S., \& {Muto}, T. 2019, \apj, 885, 52, \dodoi{10.3847/1538-4357/ab45f0}

\bibitem[{{Teague} {et~al.}(2019){Teague}, {Bae}, {Huang}, \& {Bergin}}]{teag19}
{Teague}, R., {Bae}, J., {Huang}, J., \& {Bergin}, E.~A. 2019, \apjl, 884, L56, \dodoi{10.3847/2041-8213/ab4a83}

\bibitem[{{Teague} {et~al.}(2022){Teague}, {Bae}, {Andrews}, {Benisty}, {Bergin}, {Facchini}, {Huang}, {Longarini}, \& {Wilner}}]{teag22}
{Teague}, R., {Bae}, J., {Andrews}, S.~M., {et~al.} 2022, \apj, 936, 163, \dodoi{10.3847/1538-4357/ac88ca}

\bibitem[{{Tsukagoshi} {et~al.}(2022){Tsukagoshi}, {Nomura}, {Muto}, {Kawabe}, {Kanagawa}, {Okuzumi}, {Ida}, {Walsh}, {Millar}, {Takahashi}, {Hashimoto}, {Uyama}, \& {Tamura}}]{tsuk22}
{Tsukagoshi}, T., {Nomura}, H., {Muto}, T., {et~al.} 2022, \apj, 928, 49, \dodoi{10.3847/1538-4357/ac5111}

\bibitem[{{Ueda} {et~al.}(2020){Ueda}, {Kataoka}, \& {Tsukagoshi}}]{ueda20}
{Ueda}, T., {Kataoka}, A., \& {Tsukagoshi}, T. 2020, \apj, 893, 125, \dodoi{10.3847/1538-4357/ab8223}

\bibitem[{{Vacca} \& {Sandell}(2011)}]{vacc11}
{Vacca}, W.~D., \& {Sandell}, G. 2011, \apj, 732, 8, \dodoi{10.1088/0004-637X/732/1/8}

\bibitem[{{Warren}(1984)}]{warr84}
{Warren}, S.~G. 1984, \ao, 23, 1206, \dodoi{10.1364/AO.23.001206}

\bibitem[{{Warren} \& {Brandt}(2008)}]{warr08}
{Warren}, S.~G., \& {Brandt}, R.~E. 2008, Journal of Geophysical Research (Atmospheres), 113, D14220, \dodoi{10.1029/2007JD009744}

\bibitem[{{Weingartner} \& {Draine}(2001)}]{wein01}
{Weingartner}, J.~C., \& {Draine}, B.~T. 2001, \apj, 548, 296, \dodoi{10.1086/318651}

\bibitem[{{Woitke} {et~al.}(2016){Woitke}, {Min}, {Pinte}, {Thi}, {Kamp}, {Rab}, {Anthonioz}, {Antonellini}, {Baldovin-Saavedra}, {Carmona}, {Dominik}, {Dionatos}, {Greaves}, {G{\"u}del}, {Ilee}, {Liebhart}, {M{\'e}nard}, {Rigon}, {Waters}, {Aresu}, {Meijerink}, \& {Spaans}}]{woit16}
{Woitke}, P., {Min}, M., {Pinte}, C., {et~al.} 2016, \aap, 586, A103, \dodoi{10.1051/0004-6361/201526538}

\bibitem[{{Yamamoto}(2017)}]{yama17}
{Yamamoto}, S. 2017, {Introduction to Astrochemistry: Chemical Evolution from Interstellar Clouds to Star and Planet Formation}, \dodoi{10.1007/978-4-431-54171-4}

\bibitem[{{Yoshida} {et~al.}(2022){Yoshida}, {Nomura}, {Tsukagoshi}, {Furuya}, \& {Ueda}}]{yosh22}
{Yoshida}, T.~C., {Nomura}, H., {Tsukagoshi}, T., {Furuya}, K., \& {Ueda}, T. 2022, \apjl, 937, L14, \dodoi{10.3847/2041-8213/ac903a}

\bibitem[{{Yoshida} {et~al.}(2024){Yoshida}, {Nomura}, {Furuya}, {Teague}, {Law}, {Tsukagoshi}, {Lee}, {Rab}, {{\"O}berg}, \& {Loomis}}]{yosh24}
{Yoshida}, T.~C., {Nomura}, H., {Furuya}, K., {et~al.} 2024, \apj, 966, 63, \dodoi{10.3847/1538-4357/ad2fb4}

\bibitem[{{Youdin} \& {Lithwick}(2007)}]{youd07}
{Youdin}, A.~N., \& {Lithwick}, Y. 2007, \icarus, 192, 588, \dodoi{10.1016/j.icarus.2007.07.012}

\bibitem[{{Zhang} {et~al.}(2019){Zhang}, {Bergin}, {Schwarz}, {Krijt}, \& {Ciesla}}]{zhan19}
{Zhang}, K., {Bergin}, E.~A., {Schwarz}, K., {Krijt}, S., \& {Ciesla}, F. 2019, \apj, 883, 98, \dodoi{10.3847/1538-4357/ab38b9}

\bibitem[{{Zhu} {et~al.}(2019){Zhu}, {Zhang}, {Jiang}, {Kataoka}, {Birnstiel}, {Dullemond}, {Andrews}, {Huang}, {P{\'e}rez}, {Carpenter}, {Bai}, {Wilner}, \& {Ricci}}]{zhu19}
{Zhu}, Z., {Zhang}, S., {Jiang}, Y.-F., {et~al.} 2019, \apjl, 877, L18, \dodoi{10.3847/2041-8213/ab1f8c}

\bibitem[{{Zubko} {et~al.}(1996){Zubko}, {Mennella}, {Colangeli}, \& {Bussoletti}}]{zubk96}
{Zubko}, V.~G., {Mennella}, V., {Colangeli}, L., \& {Bussoletti}, E. 1996, \mnras, 282, 1321, \dodoi{10.1093/mnras/282.4.1321}

\end{thebibliography}
\bibliographystyle{aasjournal}

%% This command is needed to show the entire author+affiliation list when
%% the collaboration and author truncation commands are used.  It has to
%% go at the end of the manuscript.
%\allauthors

%% Include this line if you are using the \added, \replaced, \deleted
%% commands to see a summary list of all changes at the end of the article.
%\listofchanges

\end{document}